%% file: cgm_flows_arxiv_v1.tex
\documentclass[useAMS,usenatbib]{mn2e}

\usepackage{times}  
\usepackage{listings}
\usepackage{graphics}
\usepackage{subfigure}
\usepackage{graphicx}
\usepackage{graphicx,xspace}
\usepackage{amssymb}
\usepackage{amsmath}
\usepackage{hyperref}
\usepackage{aas_macros}
\usepackage{lscape}
\usepackage{rotating}
\usepackage{placeins}
\usepackage{natbib}
\usepackage{color}

\newcommand{\lya}{\mbox{Ly$\alpha$}\xspace}
\newcommand{\ramses}{{\sc{ramses}}\xspace}
\newcommand{\music}{{\sc{music}}\xspace}
\newcommand{\LCDM}{{$\Lambda$\sc{CDM}}\xspace}
\newcommand{\cloudy}{{\sc{cloudy}}\xspace}
\newcommand{\eagle}{{\sc{eagle}}\xspace}
\DeclareMathOperator\erf{erf}

\title[Gas flows around simulated high-redshift galaxies]
{Gas flows in the circumgalactic medium around simulated high-redshift galaxies}

\author[P. D. Mitchell et al.]{
\newauthor Peter D. Mitchell\thanks{\rm E-mail: peter.mitchell@univ-lyon1.fr}$^{1}$,
J\'{e}r\'{e}my Blaizot$^{1}$,
Julien Devriendt$^{2,1}$,
Taysun Kimm$^{3,4}$,
\newauthor L\'{e}o Michel-Dansac$^{1}$,
Joakim Rosdahl$^{1}$ and
Adrianne Slyz$^{2}$
\\
$^{1}$Univ Lyon, Univ Lyon1, Ens de Lyon, CNRS, Centre de Recherche Astrophysique de Lyon UMR5574, F-69230, Saint-Genis-Laval, France\\
$^{2}$Sub-department of Astrophysics, University of Oxford, Keble Road, Oxford OX1 3RH, UK\\
$^{3}$Kavli Institute for Cosmology and Institute of Astronomy, Madingley Road, Cambridge CB3 0HA, UK\\
$^{4}$Department of Astronomy, Yonsei University, 50 Yonsei-ro, Seodaemun-gu, Seoul 03722, Republic of Korea\\
}

\begin{document}
\date{\today}
\pagerange{\pageref{firstpage}--\pageref{lastpage}} \pubyear{2017}
\maketitle
\label{firstpage}

\begin{abstract}

We analyse the properties of circumgalactic gas around simulated galaxies in the redshift range $z \geq 3$,
utilising a new sample of cosmological zoom simulations. These simulations are intended to be representative of the observed samples of
Lyman $\alpha$ (\lya) emitters recently obtained with the MUSE instrument (halo masses $\sim 10^{10}-10^{11} \, \mathrm{M_\odot}$). 
We show that supernova (SN) feedback has a significant impact on both the inflowing and outflowing circumgalactic
medium by driving outflows, reducing diffuse inflow rates, and by increasing the neutral fraction of inflowing gas.
By temporally stacking simulation outputs we find that significant net mass exchange occurs between inflowing and outflowing
phases: none of the phases are mass-conserving. In particular, we find that the mass in neutral outflowing hydrogen declines exponentially with radius as gas flows 
outwards from the halo centre. This is likely caused by a combination of both fountain-like cycling processes and gradual photo/collisional 
ionization of outflowing gas. 
Our simulations do not predict the presence of fast-moving neutral outflows in the CGM. Neutral outflows instead move with modest 
radial velocities ($\sim 50 \, \mathrm{kms^{-1}}$), and the majority of the kinetic energy is associated with tangential rather than radial motion.
\end{abstract}


\vspace*{-0.5cm}  

\section{Introduction}

With the advent of the MUSE instrument it has recently become possible to detect spatially extended Lyman $\alpha$ (\lya)
emission around individual, faint ($M_{\mathrm{UV}} \leq -17$), high-redshift ($3<z<6$) galaxies \cite[][]{Wisotzki16}.
This emission extends significantly beyond the rest-frame ultraviolet (UV) stellar emission of these galaxies
such that the neutral circumgalactic medium (CGM) is being probed over a significant fraction of the associated dark matter 
halo virial radii. Complementing existing pencil-beam observations of \lya in absorption \cite[e.g.][]{Peroux07,Cooper15,Prochaska15}, these observations are opening a new window 
onto the high redshift circumgalactic medium around galaxies that were too faint for study in spatially-extended emission 
with previous facilities.

Cosmological simulations are the natural tool to interpret these observations within the context of hierarchical structure formation. 
Hydrodynamical simulations predict that at this mass scale ($M_{\mathrm{H}} \sim 10^{11} \, \mathrm{M_\odot}$) and at these redshifts ($3<z<6$), 
the circumgalactic medium is highly dynamic. A hot, quasi-hydrostatic halo is not expected to have yet formed 
and filamentary flows of gas can flow directly down to the central regions of hosting dark matter halos 
\cite[e.g.][]{Birnboim03, Ocvirk08,Brooks09,Keres09}. In this regime, boundary conditions are all-important and 
therefore idealised simulations with non-cosmological initial conditions 
are arguably not a suitable tool for interpreting observations. Furthermore, both the spatial and temporal density,
temperature and velocity structure of the high-redshift ISM and CGM is predicted to be complex \cite[e.g.][]{Ceverino09,Nelson15}, 
ultimately limiting the level of confidence that can be placed in simple, idealised, analytic models, even if they are 
successful in reproducing observational phenomena \cite[e.g.][]{Verhamme08,Dijkstra12,Gronke16}.

Even with cosmological simulations, observations of \lya emission are 
particularly challenging to interpret, owing to the resonant nature of the line and the multitude of possible 
emission mechanisms \cite[e.g.][]{Fardal01,Taniguchi01,Cantalupo14,Dijkstra14}.
To accurately model \lya transfer using hydrodynamical simulations, it is first necessary to form a realistic ISM, 
including a cold phase, and with the highest spatial resolution possible \cite[][]{Verhamme12}. This is prohibitively 
expensive for large dark-matter haloes at low redshift: to resolve the thermal Jeans length across four resolution elements 
at a density, $n_{\mathrm{H}} = 10 \, \mathrm{cm^{-3}}$, and temperature, $T = 100 \, \mathrm{K}$, requires a spatial 
resolution of $13 \, \mathrm{pc}$. At present, this is barely within reach for simulations of Milky-Way mass haloes 
that reach $z=0$, even using the zoom technique \cite[][]{Hopkins14}.
Furthermore, simulating a realistic cold-phase ISM requires modelling a range of physics that is not typically included in
cosmological simulations (for example molecular cooling networks and non-equilibrium thermochemistry, radiative transfer from local sources), 
all of which act to significantly increase the computational cost of such simulations \cite[e.g.][]{Gnedin09,Wise14,Hu16,Katz16,Oppenheimer16}.

At high redshift ($z \geq 3$), the required resolution to resolve the cold-phase ISM is readily achievable at only modest expense in cosmological zoom simulations,
particularly for small haloes ($M_{\mathrm{H}} \sim 10^{11} \, \mathrm{M_\odot}$) \cite[e.g.][]{Hopkins14,Kimm15}. It
is also computationally feasible to reach the required resolution while including more complex physics (for example radiative
transfer or chemical networks) for accurately modelling the multi-phase ISM, albeit at 
greater computational expense \cite[e.g.][]{Kimm14,Katz16}. Furthermore, for small haloes ($M_{\mathrm{H}} \sim 10^{11}$)
the role of feedback from active galactic nuclei is thought to subdominant \cite[e.g.][]{Booth09,Crain15,Beckmann17,Tremmel17},
reducing the number of modelling uncertainties.
As such, the sensitivity of instruments measuring QSO absorption spectra and spatially extended 
\lya emission in this mass/redshift regime offer a unique opportunity to test state-of-the-art modelling of the ISM and CGM in
the full cosmological context.

To match observational constraints for the stellar properties of galaxies, early star formation in cosmological simulations 
must be significantly suppressed by stellar feedback processes \cite[e.g.][]{Hopkins14,Agertz15,Wang15}.
Accordingly, successful simulations predict strong mass outflows both through, and out of, haloes \cite[e.g.][]{AnglesAlcazar14,Muratov15,Wang17}. This raises
the intriguing possibility that spatially extended \lya emission may be sensitive to feedback 
\cite[in addition to the gas accretion processes that have already received significant attention in the literature, e.g.][]{Fardal01,Dijkstra09,FaucherGiguere10,Goerdt10,Rosdahl12}.
Evidence that outflows may indeed play a role in non-extended \lya detections already exists thanks to the prevalence
of red wings in \lya line profiles that are observed to be associated with blue-shifted low-ionization 
metal absorption lines in the spectra of massive Lyman-break galaxies \cite[e.g.][]{Steidel10}.
With extended \lya detections around individual objects now made possible by MUSE, it is therefore timely
to review the impact of outflows (and inflows) on the predicted properties of the neutral CGM.

Here we undertake such a study. First, we introduce a sample of $11$ high-resolution ($14 \, \mathrm{pc}$, dark matter particle mass $m_{\mathrm{DM}}=1.3 \times 10^4 \, \mathrm{M_\odot}$, stellar particle mass $m_\star^i=9.2 \times 10^3 \, \mathrm{M_\odot}$) 
cosmological zoom simulations, specifically designed to interface with the recent progress in high-redshift 
CGM observations around faint galaxies. We test these simulations against observational constraints on the 
stellar properties of high-redshift galaxies. The main part of our analysis is then dedicated to 
quantifying the predicted properties of circumgalactic gas flows and assessing the impact of SN feedback.
We place an emphasis on the properties of neutral circumgalactic hydrogen, relevant to both \lya emission 
and Lyman-limit systems.
Our intention is that the work presented here will serve as a reference point for future work involving
processing these simulations with \lya radiative transfer to enable a detailed comparison with recent observations.

Analyses similar to this study have been undertaken by a number of authors,
albeit mostly in different mass/redshift ranges, or at significantly lower resolution.
For example, \cite{VanDeVoort11} analysed inflow rates in cosmological simulations of a $34 \, \mathrm{Mpc}$ box at 
moderate resolution ($m_{\mathrm{gas}} = 1.85 \times 10^6 \, \mathrm{M_\odot}$), 
finding that (kinetic) SN feedback does not affect accretion rates onto $10^{11} \, \mathrm{M_\odot}$ haloes at $z=3$.
\cite{VanDeVoort_Cold} analysed the same simulations, finding the majority of the neutral CGM at $z=3$
is inflowing as opposed to outflowing.
\cite{FaucherGiguere11} analysed inflow rates in cosmological simulations of a $40 \, \mathrm{Mpc}$ box
at slightly lower resolution, finding that hydrodynamically decoupled winds can suppress both the net and 
inflowing accretion rates of cool gas in $10^{11} \, \mathrm{M_\odot}$ haloes at $z=3$, but only if
very high SN kinetic energy is injected.
\cite{Nelson15} analysed both inflowing and net accretion rates with a $29 \, \mathrm{Mpc}$ box at
similar resolution to \cite{VanDeVoort11}, using the calibrated subgrid models presented in \cite{Vogelsberger13}, 
which yield moderate agreement with the observed stellar mass function \cite[][]{Vogelsberger14}. They found that 
accretion rates of gas being accreted for the first time onto galaxies hosted by $10^{11.3} \, \mathrm{M_\odot}$ are reduced by a 
factor $\sim 2$ at all redshifts when feedback is included.

\cite{AnglesAlcazar14} analysed a suite of $200 \, \mathrm{pc}$ resolution (maximum) zoom simulations with
hydrodynamically decoupled, kinetic supernova-driven winds. They found that the baryon fractions within 
$10^{11} \, \mathrm{M_\odot}$ haloes at $z=2$ are reduced to roughly half the cosmological value because of SN feedback.
\cite{Brook14} analysed the baryon cycling within a suite of zoom simulations with thermal supernova feedback
(with delayed cooling) and additional thermal energy injection to model the effect of pre-SN processes associated with massive stars.
They report mass loading factors\footnote{Ratio of the mass ejection rate (by feedback) to the star formation rate.} of 
$\sim 2-3$ for haloes of mass, $8 \times 10^{11} \, \mathrm{M_\odot}$, at $z=0$ (which will evolve, on average, from
$\sim 10^{11} \, \mathrm{M_\odot}$ haloes at $z=3$), simulated with a gas particle mass $2 \times 10^5 \, \mathrm{M_\odot}$. 
\cite{Christensen16} analysed a suite of zoom simulations also featuring thermal supernova
feedback with delayed cooling. They also report similar mass-loading factors (although in this case expressed as the ratio of time-integrated outflow to integrated star formation) for similar mass haloes.
\cite{Keller16} analysed a suite of zoom simulations with gas particle mass $m_{\mathrm{gas}} = 2.2 \times 10^5 \, \mathrm{M_\odot}$.
They show that a superbubble model for SN feedback featuring a subgrid multiphase description of outflows
can also generate large mass loading factors ($\eta \sim 8$) in $10^{11} \, \mathrm{M_\odot}$ haloes.
\cite{Muratov15} \cite[see also][]{AnglesAlcazar16} performed a comprehensive analysis of outflows from very high-resolution zoom simulations featuring
thermal and momentum injection from massive stars and SN, and which have been shown to provide a good match to stellar 
constraints \cite[][]{Hopkins14}. They also report large mass loading factors ($\eta \sim 10$) for $10^{11} \, \mathrm{M_\odot}$ haloes at $z=3$,
simulated with gas particle mass, $m_{\mathrm{g}} = 5\times10^3 - 4\times10^4$.

\cite{Powell11} analysed a zoom simulation similar to the simulations presented here, but at higher resolution 
($\sim 1 \, \mathrm{pc}$) and stopping at very high redshift ($z=9$). They found that cold, filamentary accretion 
rates are roughly unaffected by SN feedback. \cite{Kimm15} analysed a suite of zoom simulations ($12 \, \mathrm{pc}$ resolution) of a single halo,
very comparable to the more massive haloes presented here (but with lower mass resolution), in this case exploring 
a range of different SN feedback models. Their mechanical feedback model is the model used in this study. 
They used an older model for star formation, with a low star formation efficiency and a simple gas density 
threshold. They report a mass-loading factor (for the mechanical feedback model) ranging from $ \sim 10$ at high
redshift to $ \sim 0.5$ at $z=3$.

The layout of this paper is as follows. We outline the technical details of our simulations,
subgrid models and post-processing analysis methodology in Section~\ref{methods}. 
In Section~\ref{validation_section}, we compare our 
simulations to constraints on the stellar properties of high-redshift galaxies inferred from 
observations. We illustrate the basic relationship between star formation and gas flows for
individual cases in Section~\ref{individ_section}. We explore the integrated baryon content
of our haloes, as well as inflow/outflow rates in Section~\ref{integrated_section}.
We explore the properties of diffuse, circumgalactic gas flows in Section~\ref{flows_sec}.
We discuss the wider implications and limitations of our results in Section~\ref{discussion_section}
and summarise our results in Section~\ref{summary_section}.

\section{Methods}
\label{methods}

\subsection{Simulations}
\label{simul_description}

\begin{table*}
\begin{center}
  \begin{tabular}{c c c c c c}
  \hline
  Halo  & $M_{\mathrm{H}}$ [Feedback] & $M_\star$ [Feedback] & $M_\star$ [No feedback] & $N_{\mathrm{DM}}$ [Feedback] & $N_{\mathrm{leaf}}$ [Feedback] \\
        & [$\mathrm{M_\odot}$]        & [$\mathrm{M_\odot}$] & [$\mathrm{M_\odot}$] & & \\
  \hline
  1  & $6.1 \times 10^9$ & $4.1 \times 10^6$ & $5.0 \times 10^8$ & $4.6 \times 10^5$ & $6.0 \times 10^5$ \\ 
  2  & $9.8 \times 10^9$ & $2.6 \times 10^7$ & $8.3 \times 10^8$ & $7.3 \times 10^5$ & $8.8 \times 10^5$ \\ 
  3  & $1.1 \times 10^{10}$ & $9.4 \times 10^7$ & $1.2 \times 10^9$ & $8.4 \times 10^5$ & $1.1 \times 10^6$ \\ 
  4  & $1.5 \times 10^{10}$ & $3.6 \times 10^7$ & $6.9 \times 10^8$ & $1.1 \times 10^6$ & $1.5 \times 10^6$ \\ 
  5  & $3.2 \times 10^{10}$ & $9.1 \times 10^8$ & $2.3 \times 10^9$ & $2.2 \times 10^6$ & $3.3 \times 10^6$ \\ 
  6  & $3.7 \times 10^{10}$ & $3.1 \times 10^8$ & $3.9 \times 10^9$ & $2.6 \times 10^6$ & $3.8 \times 10^6$ \\ 
  7  & $4.1 \times 10^{10}$ & $1.9 \times 10^8$ & $2.1 \times 10^9$ & $2.9 \times 10^6$ & $4.3 \times 10^6$ \\ 
  8  & $4.2 \times 10^{10}$ & $2.1 \times 10^8$ & $2.4 \times 10^9$ & $3.0 \times 10^6$ & $4.2 \times 10^6$ \\ 
  9  & $4.6 \times 10^{10}$ & $1.8 \times 10^9$ & $1.9 \times 10^9$ & $3.1 \times 10^6$ & $4.3 \times 10^6$ \\ 
  10 & $6.7 \times 10^{10}$ & $7.0 \times 10^8$ & $6.1 \times 10^9$ & $4.8 \times 10^6$ & $7.7 \times 10^6$ \\ 
  11 & $1.1 \times 10^{11}$ & $1.6 \times 10^9$ & $5.8 \times 10^9$ & $8.2 \times 10^6$ & $1.2 \times 10^7$ \\ 
  \hline \\
 \end{tabular}
\end{center}
\caption{Basic properties of our sample of 11 haloes at $z=3$, each of which is simulated once with
SN feedback and once without SN feedback. Properties include halo mass, $M_{\mathrm{H}}$, stellar mass, $M_\star$,
the number of dark matter particles with the virial radius, $N_{\mathrm{DM}}$, and the number of leaf gas cells
within the virial radius, $N_{\mathrm{leaf}}$.}
\label{table}
\end{table*}

The simulations presented in this paper were performed with \ramses, an Eulerian hydrodynamics 
code with adaptive mesh refinement \cite[][]{Teyssier02}. We assume a \LCDM model with cosmological 
parameters, $\Omega_{\mathrm{M}}=0.3175$, $\Omega_\lambda = 0.6825$, $\Omega_{\mathrm{B}}= 0.049$, 
$H_0 = 67.11 \, \mathrm{kms^{-1} Mpc^{-1}}$, $n_{\mathrm{s}} = 0.962$ and $\sigma_8 = 0.83$ \cite[][]{Planck14}. The 
initial conditions were generated with \music \cite[][]{Hahn11}. We first run a parent 
dark-matter-only simulation of a cosmological box to $z=3$ with $256^3$ particles over a $(20 Mpc/h)^3$ volume.
Candidate haloes are then identified and Lagrangian regions (corresponding to $150 \, \mathrm{pkpc}$ radius
spheres around these haloes at $z=3$) are resimulated at 
high resolution as a series of dark-matter-only zoom simulations (where the high-resolution region is
embedded within the lower-resolution cosmological box using the zoom technique). 
The volumes of the high resolution regions are set to $(150 \, \mathrm{pkpc})^3$ at $z=3$. 
The high-resolution regions are checked for contamination from coarse 
dark matter particles. Finally, we simulate $11$ contaminant-free haloes with full hydrodynamics
with halo masses ranging from $10^{10} \, \mathrm{M_\odot}$ to $10^{11} \, \mathrm{M_\odot}$ at 
$z=3$ (again checking for contamination). We simulate each halo both with and without 
supernova feedback. Basic information about each simulated halo is summarised in Table~\ref{table}.

The Euler equations are solved using a 2nd order Gudunov scheme. We use the HLLC Riemann solver, 
with a MinMod total variation diminishing scheme to reconstruct the intercell fluxes.
We use a Courant factor of $0.8$. 
In all of the simulations presented here we adopt the quasi-Lagrangian refinement criteria 
that cells are refined inside the zoom-region if 
$\rho_{\mathrm{DM}} \Delta x^3 + \frac{\Omega_{\mathrm{DM}}}{\Omega_{\mathrm{B}}} \rho_{\mathrm{gas}} \Delta x^3 + \frac{\Omega_{\mathrm{DM}}}{\Omega_{\mathrm{B}}} \rho_\star \Delta x^3 > 8 m_{\mathrm{DM}}^{\mathrm{HR}}$,
where $\rho_{\mathrm{DM}}$, $\rho_{\mathrm{gas}}$ and $\rho_\star$ are the respective densities
of dark matter, gas and stars, $\Delta x$ is the cell size and $m_{\mathrm{DM}}^{\mathrm{HR}}$
is the higher-resolution dark matter particle mass. 
We choose not to include additional refinement criteria based on the thermal Jeans length
\footnote{We have tested the impact of including an additional refinement criterion
that ensures that the thermal Jeans length is always resolved over $4$ leaf cells up
until the maximum refinement level of the simulation. We find this makes zero qualitative
difference to the resulting galaxy/CGM, with quantitative differences at the percent level
in galaxy stellar mass and neutral hydrogen mass in the CGM}.

The dark matter particle mass in the zoom-region is $1.3 \times 10^4 \, \mathrm{M_\odot}$. 
Star particles are formed with a minimum mass of $921 \, \mathrm{M_\odot}$.
By $z=3$, we allow up to $7$ levels of refinement beyond the initial refinement level of the 
zoom region ($12$ levels of refinement), corresponding to a leaf cell size of $14 \, \mathrm{pc}$ 
(proper) for the highest refinement level at $z=3$\footnote{
From the dark-matter-only zoom simulations, we find that dark matter
alone will lead to a maximum refinement level, $l=19$, at $z=3$ with
this scheme. In order to reach $l=21$ for the gas without the AMR
grid oversampling DM particles, we compute the density of DM (used
for computing the gravitational potential) only down
to $l=19$ (by setting {\tt \rm cic\_levelmax = 19}).
}. Note that the resolution in cosmological 
simulations is highly adaptive, such that $14 \, \mathrm{pc}$ resolution is only reached inside the ISM. 
For comparison, the typical resolution in the CGM is typically of order $\sim 100 \mathrm{pc}$
(mass-weighted average) or $\sim 500 \mathrm{pc}$ (volume-weighted average).

\subsection{Radiative heating/cooling}
\label{heating_section}

Gas heating, cooling and (H, He) ionization balance are solved assuming ionization equilibrium
for a monoatomic gas with a ratio of specific heat, $\gamma = 5/3$. Photoionization and photoheating from a uniform 
UV background \cite[][]{Haardt96} is activated after redshift, $z=8.5$. 
The UV radiation field is exponentially damped in gas cells above the
self-shielding density, $n_{\mathrm{H}} > 0.01 \, \mathrm{cm^{-3}}$.
Metal-line radiative cooling is implemented using tabulated cooling rates from \cloudy \cite[][]{Ferland98} for $T>10^4 \, \mathrm{K}$.
Below $T = 10^4 \, \mathrm{K}$, metal line cooling is implemented following \cite{Rosen95}.
In practice, this allows dense ISM gas to cool down to $10 \, \mathrm{K}$.
We do not include molecular cooling channels and as such the cold-phase ISM that forms in our simulations
is at best an approximation of a realistic cold phase. Gas in the initial conditions is assigned
a metallicity of $Z = 10^{-4} \, Z_\odot$ to help mimic the effects of molecular line cooling in unresolved low-mass
mini-haloes at high redshift. In practice, this acts to increase the amount of star formation that can occur
in low-mass satellites of our target haloes.

\begin{figure}
\includegraphics[width=20pc]{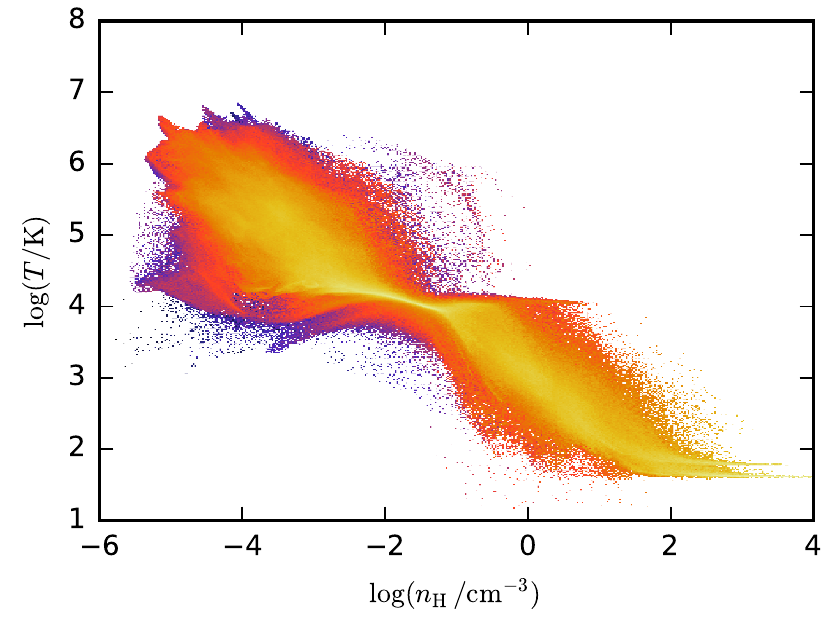}
\caption{Phase diagram of the gas within the virial radius of a $M_{\mathrm{H}} = 10^{10.5} \, \mathrm{M_\odot}$ halo at $z=3$, 
simulated with supernova feedback. Gas considered gravitationally bound to satellites has been removed.
The colour scheme is scaled logarithimically with the gas mass enclosed in each pixel, ranging from blue at low enclosed mass to 
yellow-white at high enclosed mass.
}
\label{phase_diagram}
\end{figure}

A typical $T-\rho$ diagram for a simulation with supernova feedback is shown in Fig.~\ref{phase_diagram} for a 
$M_{\mathrm{H}} = 10^{10.5} \, \mathrm{M_\odot}$ halo at $z=3$. We do not artificially pressurize gas 
that otherwise cools below a polytrope $P(\rho)$. This means that a plausibly realistic ISM density
distribution can form, which is a requirement for performing realistic \lya radiative transfer
through the ISM.

\subsection{Star formation and chemical enrichment}
\label{sf_section}

We utilise the thermo-turbulent subgrid scheme for star formation first introduced in \cite{Kimm17}
and fully described in Devriendt et al. (in preparation). In this scheme, both thermal and turbulent 
pressure support terms are balanced against gravity to determine the threshold for stability against
collapse. Gas cells that are deemed unstable then form stars following a Schmidt law, with an efficiency per
freefall time that is scaled to the local ISM conditions, following \cite{Federrath12}.

The star formation threshold relies on the computation of a thermo-turbulent jeans length, 
$\lambda_{\mathrm{J,turb}}$ \cite[][]{Federrath12}, which accounts for both thermal and 
non-thermal pressure support,

\begin{equation}
\lambda_{\mathrm{J,turb}} = \frac{\pi \sigma_{\mathrm{gas}}^2 + \sqrt{36 \pi c_{\mathrm{s}}^2 G \Delta x^2 \rho_{\mathrm{gas}} + \pi^2 \sigma_{\mathrm{gas}}^4 }}{6 G \rho_{\mathrm{gas}} \Delta x}
\label{lambda_jean}
\end{equation}

\noindent where $\sigma_{\mathrm{gas}}$ is the gas velocity dispersion measured on some physical
scale, $\Delta x$, $\rho_{\mathrm{gas}}$ is the gas density and $c_{\mathrm{s}}$ is the sound speed.
For the implementation here, $\Delta x$ is the cell size and $\sigma_{\mathrm{gas}}$ is computed
from a first order Taylor expansion of the velocity field around the centre of a given gas cell 
(Devriendt et al., in preparation).
Gas cells are considered to be unstable if $\lambda_{\mathrm{J,turb}} \leq \Delta x$. 

For cells deemed to be unstable, star formation is implemented as a Schmidt law,

\begin{equation}
\frac{\mathrm{d}\rho_\star}{\mathrm{d}t} = \epsilon_{\mathrm{ff}} \frac{\rho_{\mathrm{gas}}}{t_{\mathrm{ff}}},
\label{sfr_eqn}
\end{equation}

\noindent where $\rho_{\mathrm{gas}}$ is the gas density and $t_{\mathrm{ff}}=\sqrt{3 \pi / 32G \rho_{\mathrm{gas}}}$
is the characteristic freefall timescale \cite[][]{Schmidt59}. Stellar particles are formed using Equation~\ref{sfr_eqn} as a Poissonian process.

Contrary to a standard implementation used in cosmological simulations, the efficiency
factor, $\epsilon_{\mathrm{ff}}$ is not set to a constant, low value above some density
threshold, consistent with observations of star formation over $ \sim \mathrm{kpc}$ scales. 
Instead, the $14 \, \mathrm{pc}$ resolution of the ISM in our simulations motivates the use of a 
an efficiency per freefall time that scales with the local ISM conditions. In this case,
the local star formation efficiency per freefall time can be significantly higher than the $\sim 1\%$
efficiencies implied by observations over $\mathrm{kpc}$ scales (up to and beyond $\sim 100\%$ efficiency).

$\epsilon_{\mathrm{ff}}$ is computed by first assuming that within a given gas cell, there
is an unresolved density distribution function of lognormal form, and that there is a critical
density in this distribution above which gas can collapse to form stars. From this distribution,
a density-weighted integral of the freefall efficiency ($t_{\mathrm{ff}}^{-1}(\rho)$) is performed
for all densities, $\rho$, above the threshold, expressed here as the density contrast, $\sigma_{\mathrm{crit}}$.
$\epsilon_{\mathrm{ff}}$ is then given by normalising this integral with the the corresponding
freefall efficiency evaluated at the average density of the gas cell, $\rho_0$. Put together, this is written as

\begin{equation}
\epsilon_{\mathrm{ff}} = \frac{\epsilon_{\mathrm{acc}}}{2 \phi_{\mathrm{t}}} \exp(3 \sigma_{\mathrm{s}}^2 /8) \left[1 + \erf{\left(\frac{\sigma_{\mathrm{s}}^2-s_{\mathrm{crit}}}{\sqrt{2 \sigma_{\mathrm{s}}^2}}\right)} \right],
\end{equation}

\noindent where $\sigma_{\mathrm{s}}^2$ is the standard-deviation width of the lognormal probability distribution function,
expressed in terms of the logarithmic, dimensionless density contrast, $s \equiv \ln(\rho/\rho_0)$.
$\epsilon_{\mathrm{acc}}$ and $\phi_{\mathrm{t}}$ are dimensionless model parameters. $\epsilon_{\mathrm{acc}}$
accounts for the mass that is returned to the ISM from proto-stellar objects by jets and outflows
and is set to $0.5$, following empirical constraints \cite[see][]{Federrath12}. $\phi_{\mathrm{t}}$ accounts for the uncertainty associated with the idealised calculation
for freefall timescale of molecular clouds and is set to $0.57$, which is the best-fit value from the numerical simulations of \cite{Federrath12}.
Assuming the lognormal probability density distribution function is a physical consequence of 
isothermal, supersonic turbulence, the distribution width is given by 
$\sigma_{\mathrm{s}}^2 = \ln(1 + b^2 \mathcal{M}^2$) \cite[][]{Federrath12}.
Here, $b$ (set to $0.4$) characterises the balance of solenoidal and compressive modes for driving turbulence and
and $\mathcal{M} \equiv \sigma_{\mathrm{gas}}/c_{\mathrm{s}}$ is the sonic Mach number.

The critical logarithmic overdensity threshold for local collapse, $\sigma_{\mathrm{crit}}$, is computed
using the model of \cite{Padoan11},

\begin{equation}
s_{\mathrm{crit}} = \ln(0.067 \theta^{-2} \alpha_{\mathrm{vir}} \mathcal{M}^2),
\end{equation}

\noindent where $\theta$ is a numerical parameter calibrated to reproduce results
from the numerical simulations of \cite{Federrath12}. $\alpha_{\mathrm{vir}} \equiv 2 E_{\mathrm{k}} / |E_{\mathrm{grav}}|$
is the virial parameter, approximated here as

\begin{equation}
\alpha_{\mathrm{vir}} = \frac{5 (\sigma_{\mathrm{gas}}^2+c_{\mathrm{s}}^2)}{\pi \rho_{\mathrm{gas}} G \Delta x^2},
\end{equation}

\noindent where (as in Eqn~\ref{lambda_jean}) $\rho_{\mathrm{gas}}$ is the gas density in the cell, $\sigma_{\mathrm{gas}}$ is the gas velocity
dispersion and $\Delta x$ is the cell size.

\subsection{Supernova feedback}
\label{sn_section}

We include momentum and thermal energy injection from Type II supernovae as the sole source of feedback in 
our simulations. The scheme for supernova feedback was introduced in \cite{Kimm14} \cite[see also][]{Hopkins14,Hopkins17b}, has been described 
subsequently in \cite{Kimm15} and \cite{Kimm17}, and has been compared to other common supernova feedback 
schemes in \cite{Rosdahl17} for high-resolution, idealised disk simulations. Here, we adopt prior convention
by referring to this scheme as ``mechanical feedback''. We do not use the extensions to this scheme introduced
in \cite{Kimm17} \cite[pre-processing of the ambient ISM by photo-ionizing photons, ][]{Geen15} or
\cite{Kimm15} (temporally spaced supernova explosions and modelling porosity effects on unresolved scales).

The rationale for the mechanical feedback scheme is similar to that of the thermo-turbulent star formation model.
Namely, that when the ISM is both resolved significantly below $\mathrm{kpc}$ scales and a cold phase is allowed 
to form, it is desirable to use a feedback scheme designed to exploit both the high spatial resolution and that 
the local ISM conditions around supernova explosions are plausibly realistic \footnote{By this we mean simply that 
we have not artificially pressurized dense gas below a certain temperature.}. 
Specifically, at high enough resolution simulations can resolve the distinct stages of Sedov-Taylor SN 
explosions. If the surrounding ISM conditions are also realistic, the associated subsequent radiative losses that 
occur on resolved scales should be accurately captured. With knowledge of which stage of the SN explosion is
spatially resolved (at the time of energy/momentum injection), the radiative losses and kinetic/thermal energy 
exchanges that occur on unresolved scales can then be modelled ``subgrid'' in a physically self-consistent manner.

In the mechanical feedback scheme, momentum and thermal energy are injected into neighbouring cells 
around the host cell of a given supernova explosion. SN explosions are assumed to occur $10 \, \mathrm{Myr}$
after the formation of a given stellar particle. For each neighbouring cell, a local calculation is performed
to determine if the explosion is resolved in the energy conserving Sedov-Taylor stage or the momentum conserving
snowplow phase \cite[see][for details]{Kimm14,Kimm15}. For the former, $\Delta p$ of radial momentum is injected as

\begin{equation}
\Delta p = f_{\mathrm{c}} \sqrt{2 \chi(\Omega) m_{\mathrm{ej}} f_{\mathrm{e}} N_{\mathrm{SN}} E_{\mathrm{SN}}}
\label{momentum_adiabatic}
\end{equation}

\noindent where $N_{\mathrm{SN}}$ is the number of supernova explosions that occur inside
a given host cell, $E_{\mathrm{SN}} = 10^{51} \, \mathrm{erg}$ is the canonical energy
of individual supernova explosions, $f_{\mathrm{c}}$ is the fraction of the total outflowing mass
in the neighbouring cell, $m_{\mathrm{ej}}$ is the initial mass of the SN ejecta, $f_{\mathrm{e}}$ is a factor which ensures a smooth transition between the 
adiabatic and snowplow phases, and $\chi(\Omega)$ is the local mass-loading factor
\cite[see][for details]{Kimm14,Kimm15}.

For the case of neighbouring cells where the explosion is initially resolved in the snowplow phase, momentum is instead injected as

\begin{equation}
\Delta p = 3 \times 10^5 \, \mathrm{kms^{-1}} \, f_{\mathrm{c}} E_{\mathrm{51}}^{\frac{16}{17}} n_{\mathrm{0}}^{-\frac{2}{17}} Z'^{-0.14},
\label{snowplough}
\end{equation}

\noindent where $E_{\mathrm{51}}$ is the initial supernova energy ($N_{\mathrm{SN}} E_{\mathrm{SN}}$) 
in units of $10^{51} \, \mathrm{erg}$, $n_{\mathrm{0}}$ is the local ambient ISM hydrogen number density in units 
of $1 cm^{-3}$, and $Z'=\mathrm{max}(Z/Z_{\mathrm{\odot}},0.01)$ is the local ISM metallicity\footnote{
We use a Solar metal mass fraction of $Z_\odot = 0.02$ throughout.}. 
If the momentum injection into a given cell reduces the net kinetic energy (injected kinetic energy 
plus prexisting kinetic energy) because of vector cancelling of velocities, thermal energy is 
injected to make up the difference.

We choose model parameters such that the SN energy injected per unit stellar mass formed is
$2 \times 10^{16} \, \mathrm{erg \, g^{-1}}$. With a \cite{Chabrier03} stellar initial mass function (truncated below $0.1 \, \mathrm{M_\odot}$ and
above $100 \, \mathrm{M_\odot}$), and taking the minimum SN progenitor mass to be $8 \, \mathrm{M_\odot}$,
the energy from SN injected per unit stellar mass formed is $5.94 \times 10^{15} \, \mathrm{erg \, g^{-1}}$ (assuming
individual SN contribute $10^{51} \, \mathrm{erg}$). In practice, we find that our simulations significantly
overpredict the stellar masses of galaxies with respect to extrapolations of observational constraints
if we adopt this value. To improve this situation, we have pragmatically opted to increase the energy injected 
by SNe up to $2.0 \times 10^{16} \, \mathrm{erg \, g^{-1}}$, which represents an increase by a factor $3.4$.

There are a number of possible justifications for this energy enhancement.
Firstly, if we instead take the minimum SNe progenitor mass as $6 \, \mathrm{M_\odot}$,
our adopted parameters represent a factor $2.3$ over-injection of energy \cite[this follows from stellar models
with convective overshoot which predict that $6-8 \, \mathrm{M_\odot}$ stars explode as electron capture SNe,][]{Chiosi92}.
Secondly, our model for feedback from massive stars only includes energy from supernova explosions, neglecting
possible contributions from, for example, stellar winds, radiation, and cosmic rays. Increasing the SNe energy
by a factor two to account for this, if not an accurate way to take these processes (as we do not pre-process
the ISM before SN explode), is certainly energetically feasible \cite[e.g.][]{Agertz15}. We discuss these issues 
further in Section~\ref{missing_section}.

As well as injecting energy and momentum into the neighbouring gas cells, star particles also
release $20 \, \%$ of their mass (so $m_{\mathrm{ej}} = 0.2  m_\star^i$, where $m_\star^i$ is
the initial stellar particle mass) and inject metals with a yield of $0.05$ (as for the SN explosions,
this occurs $10 \, \mathrm{Myr}$ after the stellar particle formation time). Stellar mass loss 
and metal enrichment from stellar evolution therefore only occurs for our simulations which 
include SN feedback.

\begin{figure*}
\begin{center}
\includegraphics[width=40pc]{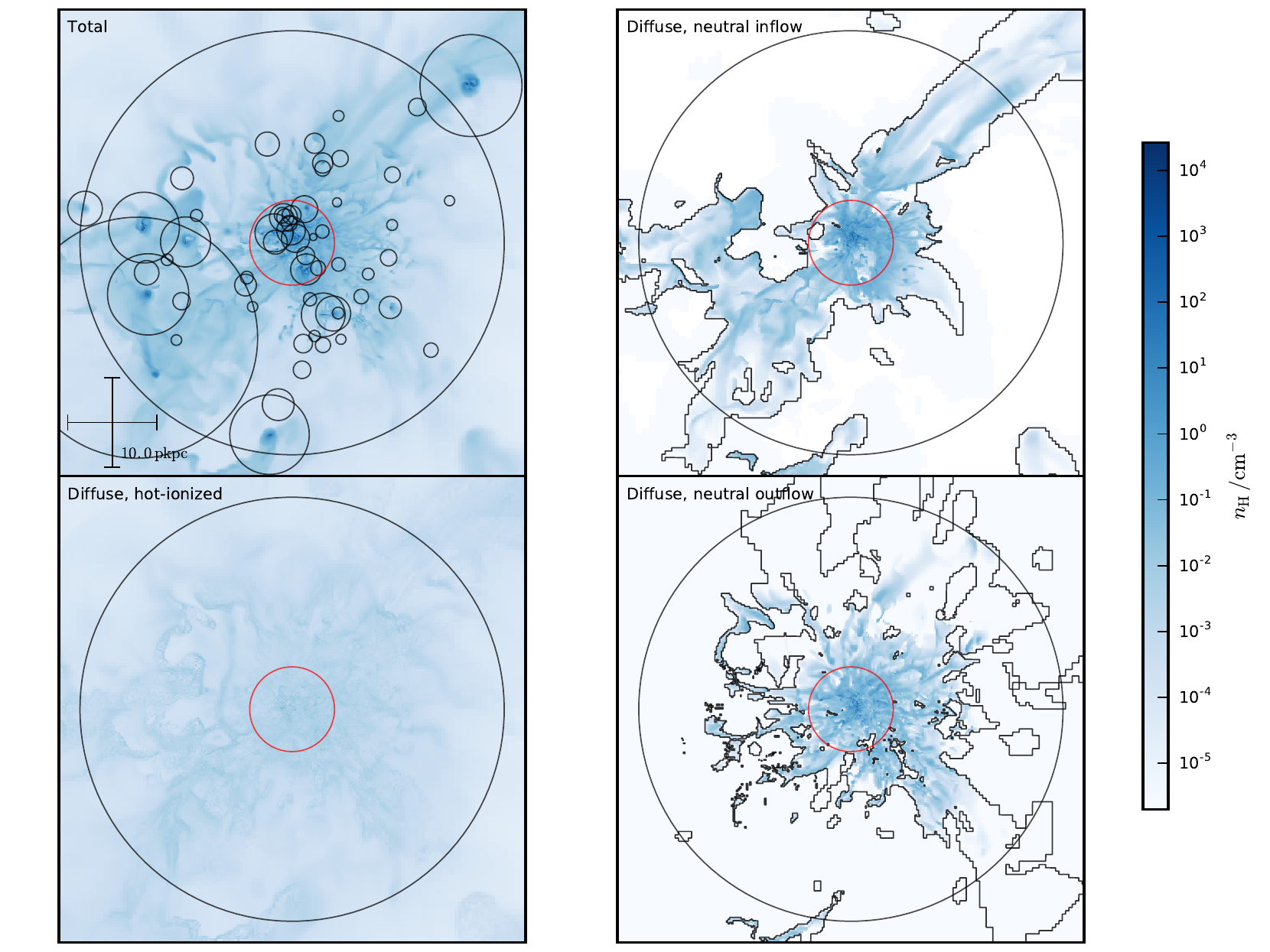}
\caption{The CGM around a $M_{\mathrm{H}} = 10^{11} \, \mathrm{M_\odot}$ halo at $z=3.9$, simulated with supernova feedback (halo 11 in \protect Table~\ref{table}). 
Each panel shows a density map for a different gas selection. 
Density maps are computed as a density-weighted average gas density along the line of sight in each pixel. 
The density map in the top-left panel includes all gas within a cube around $1.1$ of the virial radius. 
Other panels show diffuse gas (meaning gas considered gravitationally bound to satellites is removed).
These include the hot-ionized phase (bottom-left), neutral inflow (top-right) and neutral outflow (bottom-right).
For the right panels, the black contours enclose the warm-ionized phase in inflow (top-right) and outflow (bottom-right).
Large black circles indicates the virial radius, $R_{\mathrm{vir}}$. 
Red circles indicate $0.2 R_{\mathrm{vir}}$, the radius used to separate the ISM and CGM in parts of our analysis.
Small black circles mark the tidal radii of satellites. 
The black cross has a diameter of $10 \, \mathrm{kpc}$ (proper).
}
\label{density_maps}
\end{center}
\end{figure*}

\subsection{Halo identification and tree construction}

Halos are identified from the dark matter particle distribution using the AdaptaHOP algorithm \cite[][]{Aubert04},
and merger trees are constructed following \cite{Tweed09}.
In practice, the only information from the halo finder which is used in this study are the
central positions of subhaloes (which are taken to be the position of the density maxima) and the distinction
in hierarchy between central and satellite subhaloes. We measure halo masses and virial radii by computing
the spherical radius within which the mean enclosed density is $200$ times the critical density of the Universe
at a given epoch.

\subsection{Baryon compartmentalization}
\label{compartment_sec}

To analyse the CGM content of our simulated haloes, we split
the baryons enclosed within the virial radius of a given halo into
a number of components. First, we divide hydrogen between neutral and
ionized phases, using the internal \ramses calculation described in
Section~\ref{heating_section}. We further subdivide ionized hydrogen at a temperature,
$T=10^{4.5}\,\mathrm{K}$, into warm-ionized and hot-ionized phases. 
This acts to approximately separate between photo-ionized hydrogen at 
$T \sim 10^4 \, \mathrm{K}$ and collisionally ionized hydrogen at 
hotter temperatures (see the phase diagram in Fig.~\ref{phase_diagram}).
We also split each gas phase into radially inflowing and
outflowing components.

The relationship between these various gas phases is illustrated in
Fig.~\ref{density_maps}, which shows a series of projected density maps for
the most massive halo in our sample at $z=3.5$ (during an outflow
event). This halo has a mass, $M_{\mathrm{H}} = 10^{11}$, at $z=3$.
There is a clear visual distinction between the filamentary
distributions of neutral and warm-ionized inflow, the distributions of warm-ionized and  neutral outflow, and the
volume filling distribution of the hot-ionized phase.

A second step in our definition of the CGM is that unless otherwise 
stated, we exclude gas considered to be
gravitationally bound to satellites from our analysis. Specifically, 
we compute a tidal radius for each satellite subhalo, 
and subtract baryonic material enclosed within this radius. To approximate
the tidal radius, $r_{\mathrm{t}}$ we follow \cite{Binney08} and set

\begin{equation}
r_{\mathrm{t}} = R_0 \left(\frac{m}{M(R_0) \left(3 - \frac{\mathrm{d}\ln{M}}{\mathrm{d}\ln(R)}|_{R=R_0}\right)}\right)^{1/3},
\end{equation}

\noindent where $R_0$ is the distance from the satellite centre to the host 
centre, $M(R_0)$ is the mass of the host enclosed within $R_0$ and $m$ is the mass of
the satellite (the mass within $r_{\mathrm{t}}$). This relation is appropriate
for a satellite on a circular orbit within a spherically symmetric host.

Finally, in Section~\ref{CGM_section}, we compare the mass in the CGM
with different components of our simulated galaxies. Out of necessity,
we introduce a distinction between the central galaxy and the CGM by dividing at
$r = 0.2 \, R_{\mathrm{vir}}$ (indicated by the red circle
in Fig.~\ref{density_maps}). This radius approximately marks the
point below which there starts to be a significant amount of mass
in $n_{\mathrm{H}} > 1 \, \mathrm{cm^{-3}}$ gas, characteristic
of the ISM.

\subsection{Stellar luminosities}

To compare with observations, we predict the luminosities of our galaxies at different wavelengths.
To compute luminosities for stellar particles, we make use of the stellar 
population synthesis models presented in \cite{Bruzual03}. We assume a 
\cite{Chabrier03} initial mass function.
The \cite{Bruzual03} spectral models are computed for a grid of stellar ages and metallicities. We compute
spectra on a particle-by-particle basis by interpolating linearly in age and logarithmically in metallicity.
Spectra for each particle are then convolved with the desired filter bandpass to compute luminosities and
magnitudes. We do not attempt to account for the attenuation of starlight by dust and we do not include any 
additional emission mechanisms (for example, nebula line emission from HII regions) beyond those included
in the \cite{Bruzual03} models.

\subsection{Gas fluxes}

\begin{figure*}
\begin{center}
\includegraphics[width=40pc]{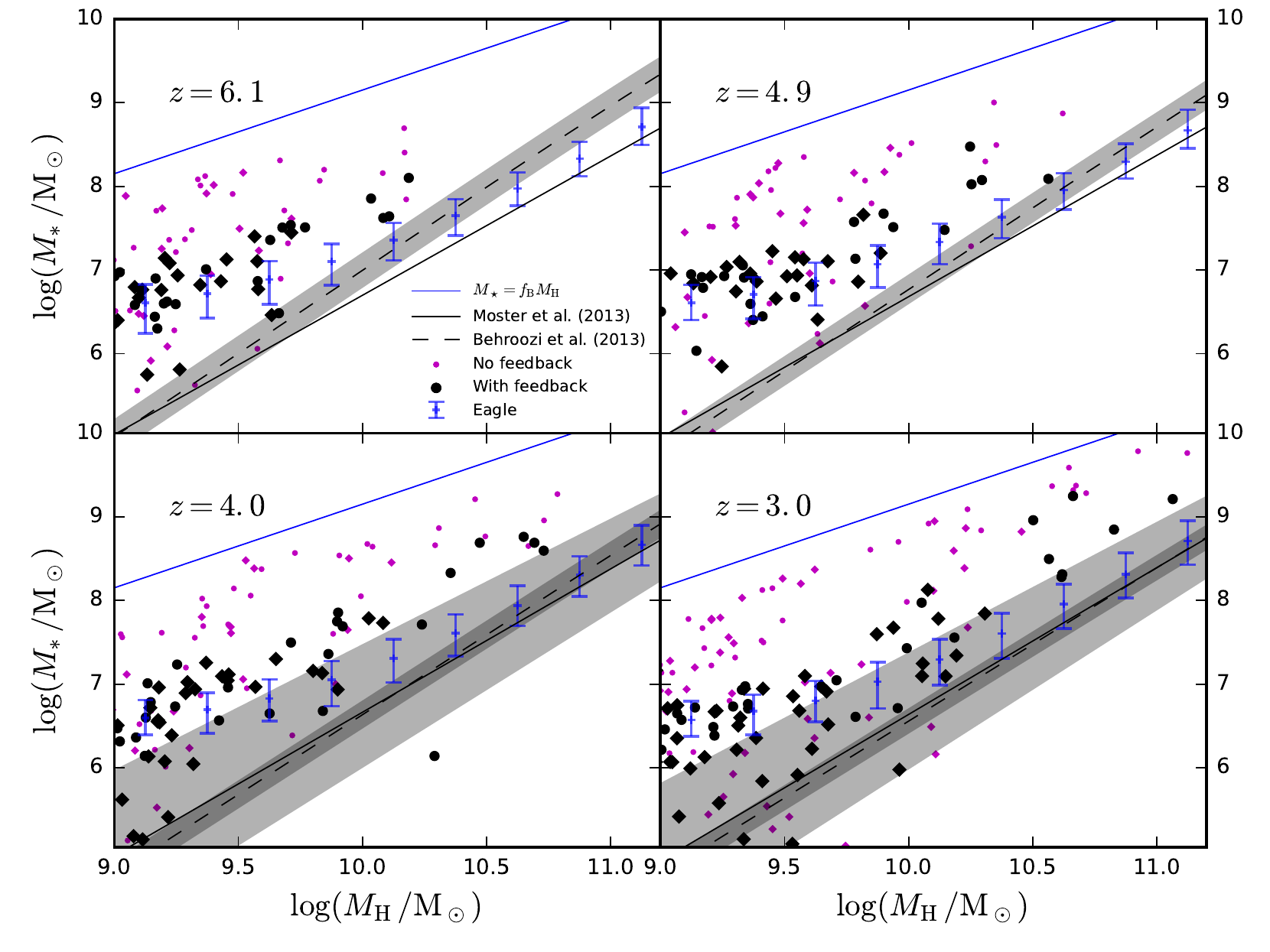}
\caption{Stellar mass as a function of halo mass for four different redshifts.
Small magenta points show galaxies simulated without SN feedback.
Large black points show galaxies simulated with SN feedback.
Circles indicate central galaxies and diamonds indicate satellites.
Stellar masses are computed by summing the stellar mass within $0.2$ of the halo virial radius.
For satellite galaxies, the halo mass plotted is the maximum past progenitor mass extracted from the merger trees.
Blue data points show the $16, 50$ and $84^{\mathrm{th}}$ percentiles of the distribution from the reference \eagle simulation.
The blue line indicates maximal conversion of baryons into stars, assuming the baryon fraction of total mass accreted through the virial radius is equal to the universal value.
The black lines \protect \cite[dashed,solid][]{Behroozi13,Moster13} show extrapolations to empirical constraints from abundance matching modelling.
Grey shaded regions show extrapolations to the quoted $1 \sigma$ uncertainties from these studies \protect \cite[which are only available for $z \leq 4$ for][]{Moster13}.
}
\label{mstar_mhalo}
\end{center}
\end{figure*}

We measure fluxes through spherical surfaces to to quantify the flows of gas through the CGM of our simulated galaxies.
Fluxes are computed separately for inflowing and outflowing components and unless
otherwise stated we remove gas considered bound to satellites from the measurements.
To compute the flux of gas at the surface of a sphere, we finely sample the surface with $N$ points\footnote{
$N$ is set to 10 multiplied by the number of gas cells expected to intersect the surface. The surface points
form a regular grid in spherical coordinates and so all surface points have equal solid angle. 
We find that the fluxes are well converged with respect to increasing $N$.}
and then compute the gas cell which encloses each point. For example, the mass flux of inflowing gas
through the surface is then given by

\begin{equation}
\dot{M} = 4 \pi R^2 \frac{1}{N} \sum_{i=1}^{N_{\mathrm{inflow}}} \rho_{\mathrm{gas},i} \, v_{\mathrm{rad},i} \, ,
\label{mass_flux_eqn}
\end{equation}

\noindent where $R$ is the radius of the surface, $\rho_{\mathrm{gas},i}$ is the density of the cell enclosing
a point, $i$, on the surface, $v_{\mathrm{rad},i}$ is the corresponding radial velocity of that cell and $N_{\mathrm{inflow}}$ 
is the number of surface points associated with inflowing cells (that are not enclosed within the tidal radii of satellites).

This approach can be generalized to compute radial profiles of various scalar quantities.
Eqn~\ref{mass_flux_eqn} represents the specific case of computing $\mathrm{d}p/\mathrm{d}r$, the
total radial momentum per unit radius across a spherical surface (which is equivalent to the 
mass flux, $\mathrm{d}m/\mathrm{d}t$). By removing $v_{\mathrm{rad},i}$ from the expression,
we can compute the radial mass profile, $\mathrm{d}m/\mathrm{d}r$. By replacing with $v_{\mathrm{rad},i}$
with specific energy (either thermal or kinetic) or metallicity, we can compute the radial
energy and metallicity profiles, $\mathrm{d}E/\mathrm{d}r$ and $\mathrm{d}Z/\mathrm{d}r$.

\section{Model validation}
\label{validation_section}

\begin{figure*}
\begin{center}
\includegraphics[width=40pc]{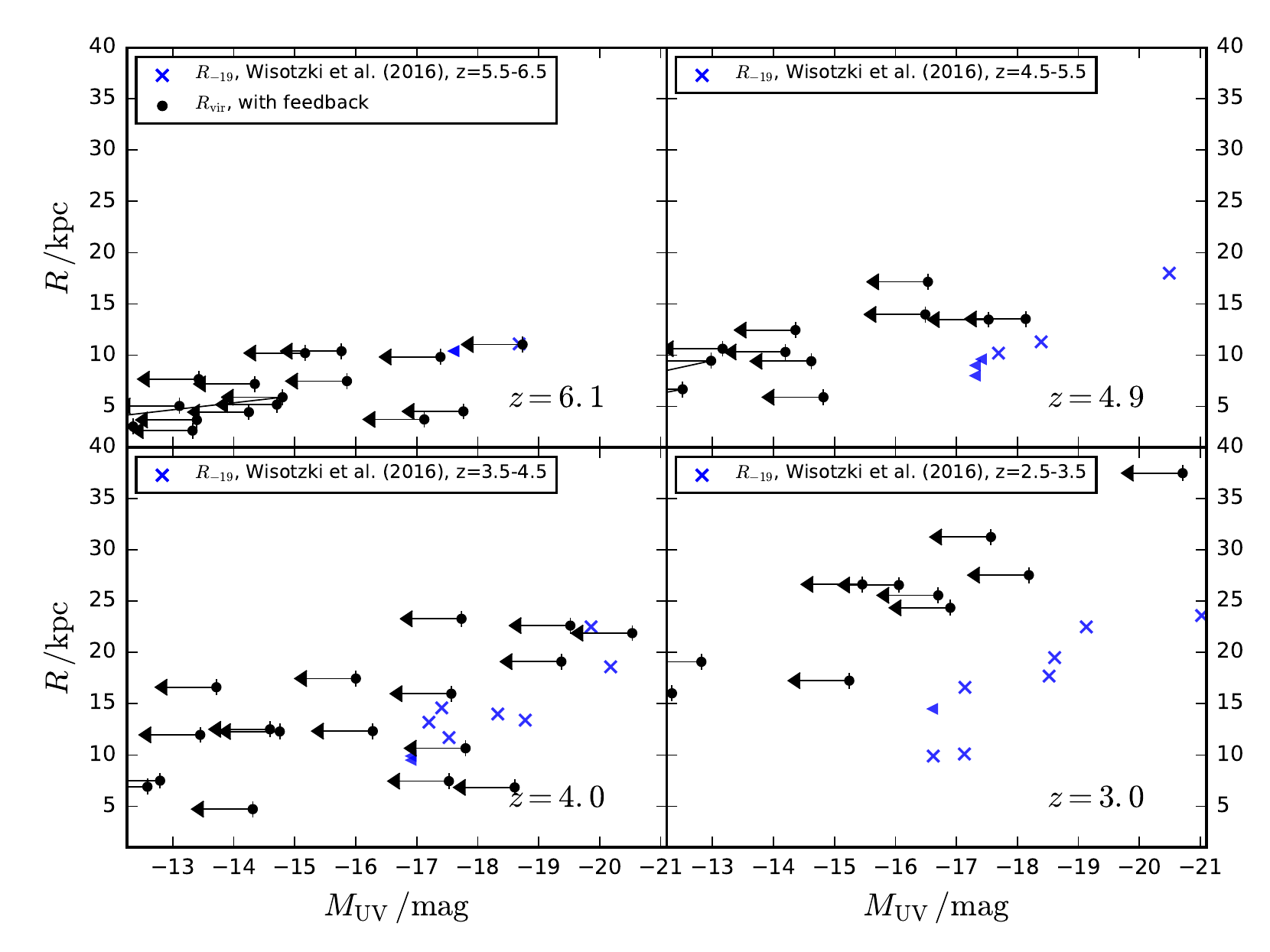}
\caption{The neutral CGM scales traced by extended \lya emission.
Black points show simulated halo virial radii for central galaxies as a function of rest-frame $1500 \sim \mathrm{\AA}$ ultraviolet absolute magnitude (not including any correction for dust).
Halo virial radii are shown to act as a reference point for the physical scales associated with the analysis presented in this study.
They are to be compared to isophotal radii, $R_{\mathrm{-19}}$, (blue crosses) that indicate the limiting sensitivity of spatially extended \lya emission in the \protect \cite{Wisotzki16} observational sample.
From this comparison we see that \lya emission traces gas over a significant fraction of the virial radius.
Blue triangles indicate upper limits to the ultraviolet luminosity for galaxies in the \protect \cite{Wisotzki16} that are undetected with Hubble Space Telescope.
The observed data are not corrected for dust attenuation.
To serve as a visual guide, black arrows indicate the typical $1$ magnitude dust attenuation estimated for galaxies at these magnitudes/redshifts \protect \cite[][]{Bouwens16}.
Each panel corresponds to a different redshift, as labelled. 
Observed \lya emitters have been binned in redshift accordingly, as labelled.
Note that because of sensitivity limits and surface brightness dimming, $R_{\mathrm{-19}}$ should not be interpreted as the intrinsic size of \lya haloes.
}
\label{muv_rhalo}
\end{center}
\end{figure*}

To validate our numerical simulations as a tool for studying the CGM, we first establish
the extent to which they provide a reasonable description of faint, high-redshift galaxies.
To do so, we compare to constraints inferred from observations for galaxy stellar masses, star formation 
rates and sizes.

In Fig.~\ref{mstar_mhalo}, the stellar masses of our simulated galaxies are plotted as a function of halo mass.
The distribution of points can be compared to the black lines, which show empirical constraints from
abundance matching \cite[][]{Behroozi13,Moster13}. This comparison represents the zeroth order constraint for samples of cosmological
zoom simulations, for which it is not possible to predict the stellar mass function.
The abundance matching studies shown here are
constrained at $z \geq 3$ by observational datasets which are insufficiently deep to probe the
stellar mass range shown here ($M_\star < 10^9 \, \mathrm{M_\odot}$), and as such the relations shown represent extrapolations
(both from higher mass galaxies at $z \leq 4$ and from lower mass galaxies at lower redshift).
 
When feedback is included (black points), our simulations appear in reasonable agreement with
the extrapolated empirical constraints at $z=3$ for the low-mass haloes in our sample ($\log(M_{\mathrm{H}}/\mathrm{M_\odot}) < 10.5$).
On the other hand, it appears that the simulated sample has stellar masses that are too slightly high\footnote{
Despite the fact that we over-inject SN energy by a factor $\approx 2-3$, see Section~\ref{sn_section}.}.  
over the halo mass range, $\log(M_{\mathrm{H}}/\mathrm{M_\odot}) > 10.5$.
Compared to the extrapolated best-fit abundance matching estimates, our simulated galaxies in this mass range 
have stellar masses that a factor $7-8$ too large on average, although the scatter at fixed halo mass is significant. 
Compared to the no-feedback case, including SN feedback for these haloes reduces 
stellar masses by roughly a factor $7$ (although notably Halo 9 from Table~\ref{table} has almost the same
stellar mass with and without feedback). Including feedback also reduces halo masses \cite[see][]{Schaller15}, which is not accounted for in abundance
matching and therefore artificially increases the level of disagreement by a small amount. 

Given that the abundance matching relations represent extrapolations in this mass range at these redshifts, 
we choose to also show results from the reference \eagle simulation \cite[][blue points]{Schaye15}. The
reference \eagle simulation is a large volume simulation 
($100^3 \, \mathrm{Mpc}^3$) which was explicitly calibrated to 
reproduce the local stellar mass function and has been shown to be consistent with observational constraints
on the high-redshift stellar mass function \cite[][]{Furlong15}. While observational 
constraints in the relevant mass/redshift range for this study are not available, \eagle
does in effect provide a different way to extrapolate the observed high-redshift stellar mass functions
down to lower stellar masses, acting as a consistency check with the
abundance matching estimates. \eagle is indeed consistent with the abundance matching extrapolations
in the halo mass range where galaxies are well resolved (at approximately 
$M_{\mathrm{H}} \geq 10^{10} \, \mathrm{M_\odot}$), despite halo masses being reduced
by baryonic effects such as feedback \cite[][]{Schaller15}. We note also that in the mass range shown, 
the distributions from \eagle are unaffected by AGN feedback\footnote{We have checked this by comparing 
to an \eagle simulation run without AGN feedback}.
On balance, the combined picture from \eagle and abundance matching suggests that
it is likely that feedback from stars is too weak in our simulations in the 
$\log(M_{\mathrm{H}}/\mathrm{M_\odot}) > 10.5$ range, although the extrapolation
involved means that it is not possible to form a definitive conclusion.

We have also compared our simulations to observational constraints on galaxy star formation rates
and galaxy sizes (not shown). From these comparisons we find no obvious tension with the
observational data if we extrapolate the observed trends. As with the stellar masses, the data are 
insufficiently deep to be truly constraining.

\subsection{Detectability and the $\mathrm{Ly}\alpha$-CGM connection}

As well as assessing the realism of simulated galaxies, another simple question is whether our simulated 
galaxies \cite[intended to be representative of the][\lya-emitter sample]{Wisotzki16} would be detectable 
in the rest-frame ultraviolet in a deep field observed with Hubble Space Telescope (HST).
This question is answered in Fig.~\ref{muv_rhalo}, which shows that the brighter galaxies that we simulate 
are indeed comparable in terms of ultraviolet luminosities to the \cite{Wisotzki16} sample. 
While we also simulate galaxies that are fainter than those detected
in \cite{Wisotzki16}, we note that spatially extended \lya is detected around \lya emitters that are
formally undetected in the rest-frame ultraviolet by HST. The simulated galaxies that are fainter
than this sensitivity limit may therefore still be of interest in future work where we will compare the simulations
and observations in terms of spatially extended \lya emission. We do not attempt to model attenuation of 
ultraviolet photons by dust, and so the ultraviolet luminosities of our simulated galaxies will be overestimated,
likely by $1$ magnitude or less \cite[][]{Bouwens16}.

Fig.~\ref{muv_rhalo} also demonstrates that the extended \lya emission observed by \cite{Wisotzki16}
likely traces a significant fraction of the halo virial radius, supporting the interpretation that 
these observations trace the neutral CGM around high-redshift galaxies (see also Leclercq et al., in preparation
for a similar analysis with a larger sample). Incidentally, given that
the detected emission is always enclosed within the associated halo virial radii, the choice
to truncate our analysis of the CGM at the halo virial radius is reasonable within the context of 
interpreting extended \lya emission.

\section{Flows of gas in single objects}
\label{individ_section}

\begin{figure*}
\begin{center}
\includegraphics[width=40pc]{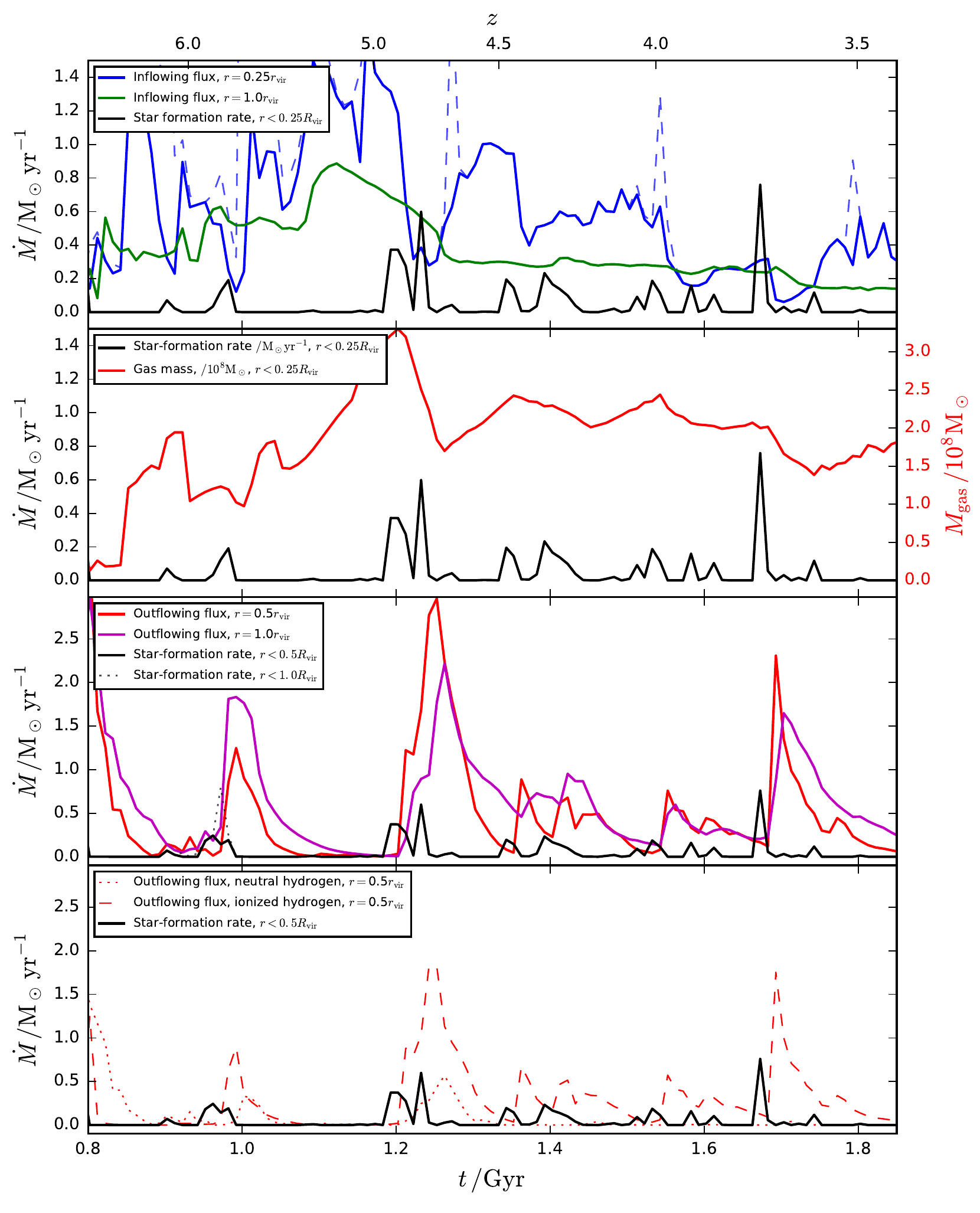}
\caption{Star formation and flux histories of a simulated halo of mass, $\log(M_{\mathrm{H}}/\mathrm{M_\odot}) = 10.1$ at $z=3$ \protect (halo 3 in Table~\ref{table}).
{\it Top:} Flux of inflowing gas through the virial radius (green) and through a surface at $0.25$ of the virial radius (solid blue).
A dashed blue line shows the flux of inflowing gas through the $0.25$ surface, in this case including gas that is gravitationally bound to satellites.
Also shown is the star formation rate (black) for stars forming within $0.25$ of the virial radius.
{\it Second:} Star formation rate (black) for stars forming within $0.25$ of the virial radius is compared to the total gas mass within the same radius (red).
Note that the units of the two lines are different (see left and right axes labels).
{\it Third:} Flux of outflowing gas through the virial radius (magenta) and through a surface at $0.5$ of the virial radius (red).
Also shown is the star formation rate for stars forming within $0.5$ of the virial radius (solid black), and within the virial radius (dotted black).
{\it Bottom:} Flux of outflowing neutral (dashed red) and ionized (dotted red) hydrogen through a surface at $0.5$ of the virial radius.
Also shown is the star formation rate for stars forming within $0.5$ of the virial radius (black line).
For all panels, the star formation rates shown are computed over a $10 \, \mathrm{Myr}$ timescale.
}
\label{sfh_illustration}
\end{center}
\end{figure*}

\begin{figure*}
\begin{center}
\includegraphics[width=40pc]{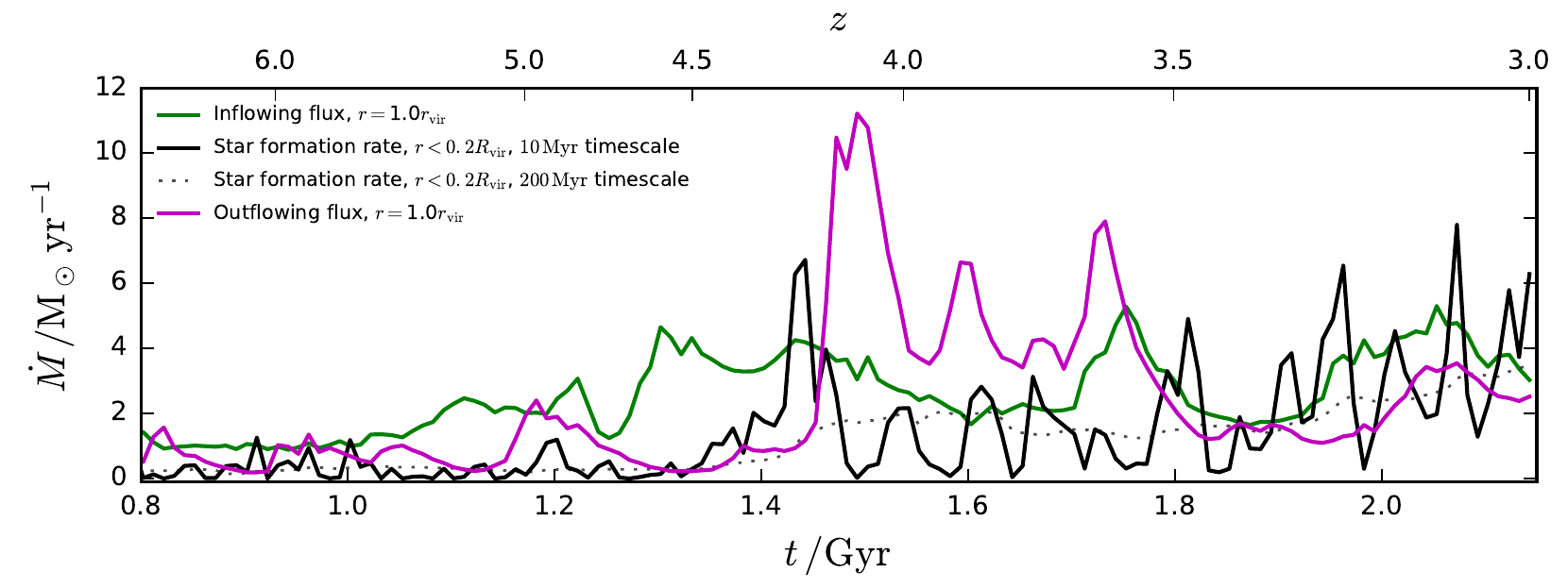}
\caption{Star formation and flux histories of a simulated halo of mass, $\log(M_{\mathrm{H}}/\mathrm{M_\odot}) = 11.1$ at $z=3$ \protect (halo 11 in Table~\ref{table}).
Fluxes of inflowing (green) and outflowing (magenta) gas through the virial radius are shown.
Also shown is the star formation rate for stars forming within $0.2$ of the virial radius, computed over a $10 \, \mathrm{Myr}$ (solid black) and a $200 \, \mathrm{Myr}$ (dotted black) timescale.
}
\label{sfh_illustration2}
\end{center}
\end{figure*}

Before presenting an analysis of the entire sample, it is illustrative to first consider single objects.
The lower-mass haloes from our sample are particularly well suited for this purpose because individual 
episodes of star formation and mass outflows are cleanly separated in time, highlighting
the relationship between star formation and outflows. 

Fig.~\ref{sfh_illustration} shows the star formation history, as well as the inflowing and outflowing fluxes of gas for a
halo with a final mass of $\log(M_{\mathrm{H}}/\mathrm{M_\odot}) = 10.1$ at $z=3$ (halo 3 in Table~\ref{table}). The top panel illustrates the temporal
relationship between inflow and star formation. Inflowing gas crosses the virial radius (green line) and a comparable
mass also crosses an inner surface at $0.25$ of the virial radius (solid blue line). Inflow through the inner surface (which 
can be regarded as the  accretion rate of the galaxy) is less steady than the inflow into the halo. From visual inspection,
the spikes seen in the inflowing flux through the inner surface are related to galaxy mergers (the corresponding fluxes
when material that is gravitationally bound to satellites is included are shown by the dashed blue line).

The star formation history in this halo is very bursty, with individual bursts typically occurring over a duration of $\sim 50 \, \mathrm{Myr}$.
Comparing the star formation history with the inflow through $0.25 R_{\mathrm{vir}}$, it is apparent that the vast
majority of the inflowing gas is not converted into stars. It is also apparent that some of main bursts of star formation
are preceded by a significant positive fluctuation in the inflow rate, but others are not. This serves to underline that there are local processes within the 
simulated central galaxy besides the gas accretion rate that control the star formation rate.

A different perspective is shown by the second panel of Fig.~\ref{sfh_illustration}, which compares the star formation
history (in units of mass rate) to the gas mass within $0.25 R_{\mathrm{vir}}$. 
The twin bursts of star formation that occur just after $z=5$ are associated with a large increase in the total gas mass.
These twin bursts are then followed by a significant reduction in the ISM gas mass. Later bursts also affect the
growth of the ISM mass in a similar manner.

The third panel of Fig.~\ref{sfh_illustration} illustrates the temporal relationship between star formation and mass
outflows. In this panel, we show the star formation taking place within two spherical apertures and the corresponding
outflowing mass flux of  gas at those radii. There is a clear and simple relationship between star
formation episodes and strong mass outflow events. Star formation takes place and (after a small but finite delay) is 
followed by outflow events characterised by an exponentially-decaying temporal profile. The time delay can be partly 
attributed to our modelling of mechanical Type II SN feedback, where the time delay between stars forming and 
SN explosions is $10 \, \mathrm{Myr}$. From the area under the star formation and outflow fluxes, it
is readily apparent that the mass ejected from the galaxy is significantly larger than the mass turned into stars. At
$z=3$, the ratio of time-integrated outflow to star formation is approximately ten. Correspondingly, by comparing to the 
second panel, it can be inferred that the reduction in gas mass that occurs after star formation bursts is primarily because of outflows, rather than from star formation.
The third panel of Fig.~\ref{sfh_illustration} also illustrates the role of satellites in driving outflows.
The star formation that occurs beyond $0.25 \, R_{\mathrm{vir}}$ is associated with satellite galaxies (this occurs
once just after $z=6$). This
star formation then drives outflows out through the surfaces at $0.5, 1 \, R_{\mathrm{vir}}$.

The bottom panel of Fig.~\ref{sfh_illustration} decomposes the flux shown in the third panel between neutral and ionized
phases of outflowing hydrogen (note helium is excluded hence the lower normalization compared to the upper panels). This 
illustrates that the majority of hydrogen mass in the outflow is ionized, but that there is also a residual amount of mass 
in neutral hydrogen. The importance of neutral outflows does however increase with increasing redshift.
Neutral outflow events at a given surface in the halo lag behind ionized outflows, forming a a clear temporal sequence.

While the relative simplicity of the simulated halo shown in Fig.~\ref{sfh_illustration} makes it ideally suited to illustrate
the temporal relationship between inflow, star formation, and outflow, it is also useful to briefly consider a more massive
halo. Fig.~\ref{sfh_illustration2} shows the most massive halo in our sample (halo 11 in Table~\ref{table}), which has a final mass of 
$\log(M_{\mathrm{H}}/\mathrm{M_\odot}) = 11.1$ at $z=3$. 
Similar to the smaller halo shown in Fig.~\ref{sfh_illustration}, this halo also exhibits strong short term fluctuations (tens of $\mathrm{Myr}$) 
in the star formation rate. The star formation rate computed over a $200 \, \mathrm{Myr}$ timescale (as traced by ultraviolet
starlight) is comparatively smooth in time such that it would be consistently detected in the UV, unlike the smaller halo. 
While there are still strong  mass outflows that leave the halo (magenta line), it can 
be seen qualitatively that the ratio of integrated outflow to integrated star formation is much smaller in the massive halo 
(the ratio is of order unity integrated to $z=3$).

\section{Halo baryon fractions and SN feedback}
\label{integrated_section}

\begin{figure}
\includegraphics[width=20pc]{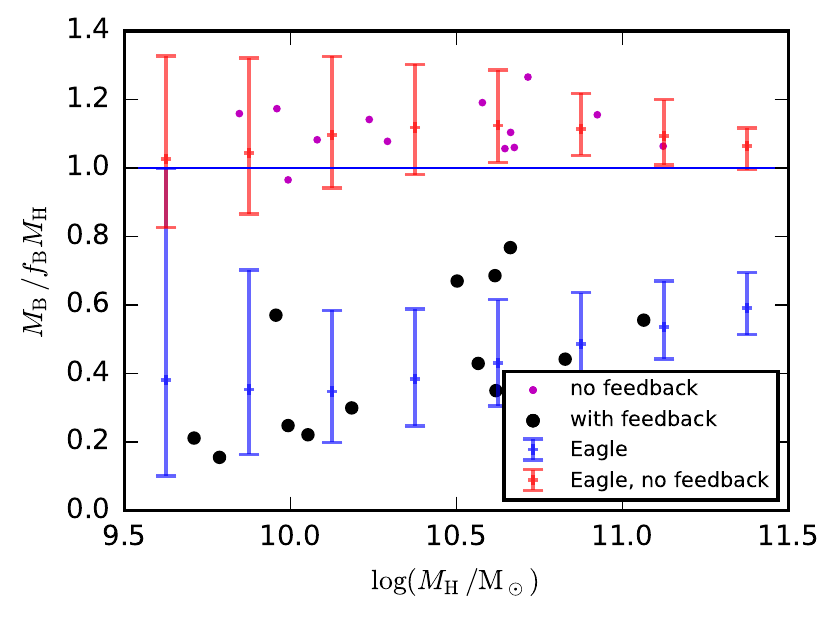}
\caption{Total baryon mass within the virial radius, $M_{\mathrm{B}}$, of simulated central haloes, plotted as a function of halo mass at $z=3$.
Here, baryons considered to be gravitationally bound to satellites are included in $M_{\mathrm{B}}$.
Baryon masses are normalized by the halo mass, $M_{\mathrm{H}}$, scaled by the universal baryon fraction, $f_{\mathrm{B}}$.
Black points show simulated haloes in runs that include SN feedback.
Magenta points show simulated haloes in runs that do not include SN feedback.
Blue points show the $10$, $50$ and $90^{\mathrm{th}}$ percentiles of the distribution in the reference \eagle simulation at $z=3$, including only central objects.
Red points show the corresponding distribution for an \eagle simulation without feedback.
The blue horizontal line marks an enclosed baryon fraction of unity, where the baryon fraction inside the halo is equal to the universal value.
For both Eagle and our simulated haloes, the halo virial radius is defined as the radius at which the enclosed average density is equal to $200$ times the critical density of the universe. 
}
\label{fbaryon}
\end{figure}

Fig.~\ref{fbaryon} shows the baryon fractions of our simulated haloes at $z=3$.
Without feedback (small magenta points), the baryon fractions are typically above
the universal value \cite[see the discussion in footnote 6 of][]{Kimm15}.
When supernova feedback is included (black points), the baryon fractions are significantly
lowered to below the universal value. The reduction in baryon fractions associated with 
introducing supernova feedback can be attributed to a combination of two effects. In part
gas is ejected from haloes as an outflow and it part the accretion rate of inflowing material is lowered.
Most of the haloes are consistent with the distribution of baryon fractions in the \eagle 
simulation, indicating that the net efficiency of feedback in removing baryons from haloes 
is similar. Given that \eagle produces lower
stellar masses (which are better in agreement with extrapolations to abundance matching
constraints) than our simulations for $M_{\mathrm{H}} > 10^{10.5} \, \mathrm{M_\odot}$,
this underlines that there are more factors at work which determine the efficiency of star 
formation than just the mass of baryons within the virial radius.

In the following subsections, we investigate the origin of the reduced halo baryon fractions.
In Section~\ref{outflow_subsec}, we measure the mass removed by outflows. In Section~\ref{inflow_sne_sec}, 
we show that feedback also reduces the accreted mass. In Section~\ref{CGM_section}, we show that
the decrease in halo baryon fractions corresponds primarily to a reduction of mass in galaxies
and has only a small effect on the total mass in the CGM. We then explore the CGM
mass budget split into outflowing/inflowing neutral/ionized components.

\subsection{Outflow masses and loading factors}
\label{outflow_subsec}

The relationship between star formation and mass outflows is demonstrated anecdotally in 
Fig.~\ref{sfh_illustration} and Fig.~\ref{sfh_illustration2}. To quantify this effect for
our full sample, we introduce an integrated loading factor,

\begin{equation}
\eta_{\mathrm{int}}(r,t) \equiv \frac{\int_0^t \dot{M}_{\mathrm{out}}(r(t'),t') \,\mathrm{d}t'}{\int_0^t \dot{M}_\star(<r(t'),t') \,\mathrm{d}t'} ,
\end{equation}

\noindent where $\dot{M}_\star(<r(t'),t')$ is the star formation rate interior to a radius, $r(t')$, 
at a time $t'$, and $\dot{M}_{\mathrm{out}}(r(t'),t')$ is the mass flux of 
radially outflowing gas passing through a surface\footnote{In practice we always place the 
surface at a fraction of the halo virial radius, for example at $r(t') = 0.5 R_{\mathrm{vir}}(t')$}
at a radius, $r(t')$, at time, $t'$.
Hence, $\eta_{\mathrm{int}}(r,t)$ represents the ratio of integrated mass outflow to 
integrated star formation and as such, can be considered as an (integrated) efficiency 
of feedback. We prefer to use this quantity when assessing the efficiency of feedback
rather than an instantaneous mass loading factor 
($\eta_{\mathrm{ml}} \equiv \dot{M}_{\mathrm{out}} / \dot{M}_\star$) to reduce the systematics and noise 
associated with variability and phase offsets between star formation and outflow, as 
seen in Fig.~\ref{sfh_illustration} and Fig.~\ref{sfh_illustration2}. To give an idea
as to the importance of these effects, and to give a quantity which can be readily
compared to observations, we also consider the instantaneous mass loading
factor, $\eta_{\mathrm{ml}}$, with $\dot{M}_\star$ computed over a $200 \, \mathrm{Myr}$
timescale (so as to be comparable to star formation rates estimated from the UV continuum).

We also consider the outflow mass, defining

\begin{equation}
M_{\mathrm{out}}(r,t) \equiv \int_0^t \dot{M}_{\mathrm{out}}(r(t'),t') \,\mathrm{d}t' ,
\label{mout}
\end{equation}

\noindent as the integrated mass in  gas that passes radially outwards through a surface at $r(t')$.

\begin{figure}
\includegraphics[width=20pc]{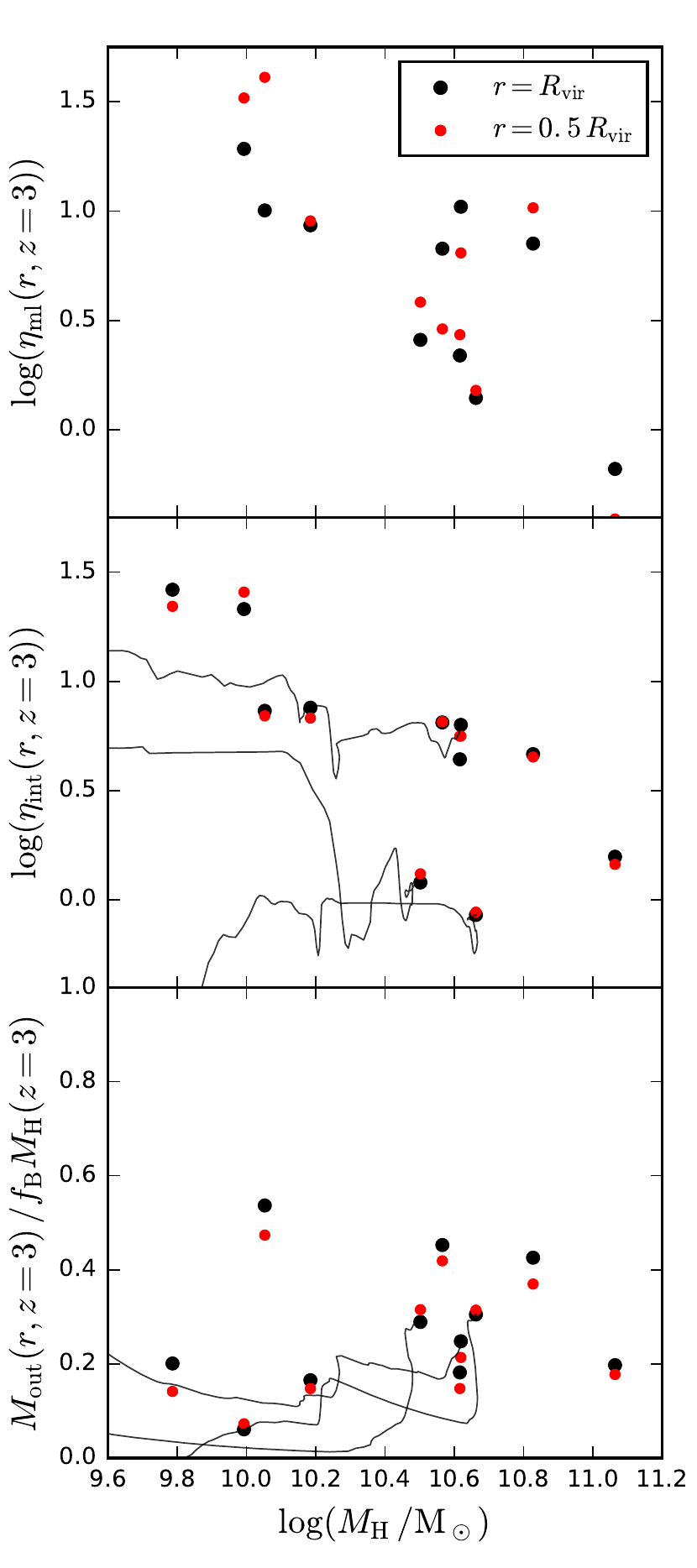}
\caption{
{\it Top:} Instantaneous mass loading factor, $\eta_{\mathrm{ml}}(r,t)$, plotted as a function of halo mass for the haloes simulated with SN feedback.
Only the central main-target haloes for the $11$ zoom simulations are shown.
Star formation rates used for this quantity are computed over a $200 \, \mathrm{Myr}$ timescale.
Red points show the loading factor for a surface placed at half the halo virial radius.
Black points show the loading factor for a surface placed at the halo virial radius.
{\it Middle:} Same as the top panel but showing instead the time-integrated loading factor, $\eta_{\mathrm{int}}(r,t)$.
The loading factors are computed by integrating down to $z=3$.
Black lines show past tracks for three example objects, with the time integral performed up to the time that a halo has a given mass.
{\it Bottom:} Same as the top panel but instead plotting the normalized, time-integrated outflow mass.
The normalization factor is the baryonic mass at the halo would have at $z=3$ if it contained the universal baryon fraction.
For the past tracks in this panel (black lines), the halo mass used is the instantaneous halo mass.
}
\label{loading_factors}
\end{figure}

In Fig.~\ref{loading_factors} we plot the instantaneous (top) and time-integrated loading factors (middle panel) of our targeted, 
simulated haloes as a function of halo mass for two different surfaces at $0.5, 1 \, R_{\mathrm{Vir}}$. 
As expected, there is significant scatter in the distribution of instantaneous mass loading factors.
The scatter is smaller if we instead consider the time-integrated loading factors.
In this case, the loading factors are virtually identical between the two surfaces (red and black points), meaning that mass outflows
driven by SN propagate all the way through, and out of, the halo (regardless of the halo mass). 
The absolute value of the integrated loading factor scales with halo mass, ranging
from values of $\eta_{\mathrm{int}} \sim 15$ at $M_{\mathrm{H}} = 10^{10} \, \mathrm{M_\odot}$ to $\eta_{\mathrm{int}} \sim 1.5$ at $M_{\mathrm{H}} = 10^{11} \, \mathrm{M_\odot}$.

The amount of outflowing mass can be seen in an absolute sense by considering the bottom panel of Fig.~\ref{loading_factors},
which instead shows the integrated outflowing masses, $M_{\mathrm{out}}$, normalized by the final 
halo mass at $z=3$. Here we see that relative to the expected baryonic growth of the halo 
($f_{\mathrm{B}} M_{\mathrm{H}}(z=3)$), the outflows expel somewhere between $20-60 \%$ of the
expected baryonic mass out of the halo. Normalized in this way, there appears to be no trend with halo mass.
This indicates that the mass dependence of the  baryon fractions shown in Figure~\ref{fbaryon} is caused
by an effect other than outflows removing mass from haloes.
Comparing the time-integrated outflow rates shown in the lower panel of Fig.~\ref{loading_factors} to the halo 
baryon fractions shown in Fig.~\ref{fbaryon}, it is apparent that outflows alone do not account for the low
baryon fractions seen when SN feedback is included: we find that outflows from the central galaxies
contribute to between $\sim 20 \, \%$ and $\sim 60 \, \%$ of the reduction in baryon fractions\footnote{This
can be seen be cross-referencing Fig.~\ref{fbaryon} with Fig.~\ref{loading_factors}.}. 

Notably, there are two simulated haloes (the third and seventh most massive) that appear to have unusually 
low integrated loading factors compared to the trend outlined by the rest of the sample (these are haloes
5 and 9 in Table~\ref{table}). 
These are (not coincidentally) the two haloes that have the highest ratio of $M_\star / M_{\mathrm{H}}$
in Fig.~\ref{mstar_mhalo}. We see in the bottom panel of Fig.~\ref{loading_factors} that these haloes
do not have unusually low integrated outflow masses relative to the rest of the sample. This indicates that
these outliers in the integrated loading factor distribution are caused by unusually efficient star formation.
The two galaxies are not unusual in their stellar specific angular momentum or half-mass radii with respect to the rest of sample
so this difference does not appear to be caused by above average compactness. The black lines in Fig.~\ref{loading_factors} show 
the past tracks for these two outlying haloes (as well as a third, non-outlier halo as a reference). The more massive
outlier (halo 9, also the halo for which feedback has almost no effect on the stellar mass) has a persistently low 
time-integrated loading factor (middle panel) and experiences significant outflows only relatively late in the haloes growth history (bottom panel).
The less massive outlier (halo 5) undergoes very little early star formation (and hence outflows), with most
of the stars formed in two large bursts around $z=4$, leading again to an abrupt increase in outflowing
mass (after which the halo does not grow significantly in mass, bottom panel).

\subsection{Inflows and SN feedback}
\label{inflow_sne_sec}

\begin{figure}
\includegraphics[width=20pc]{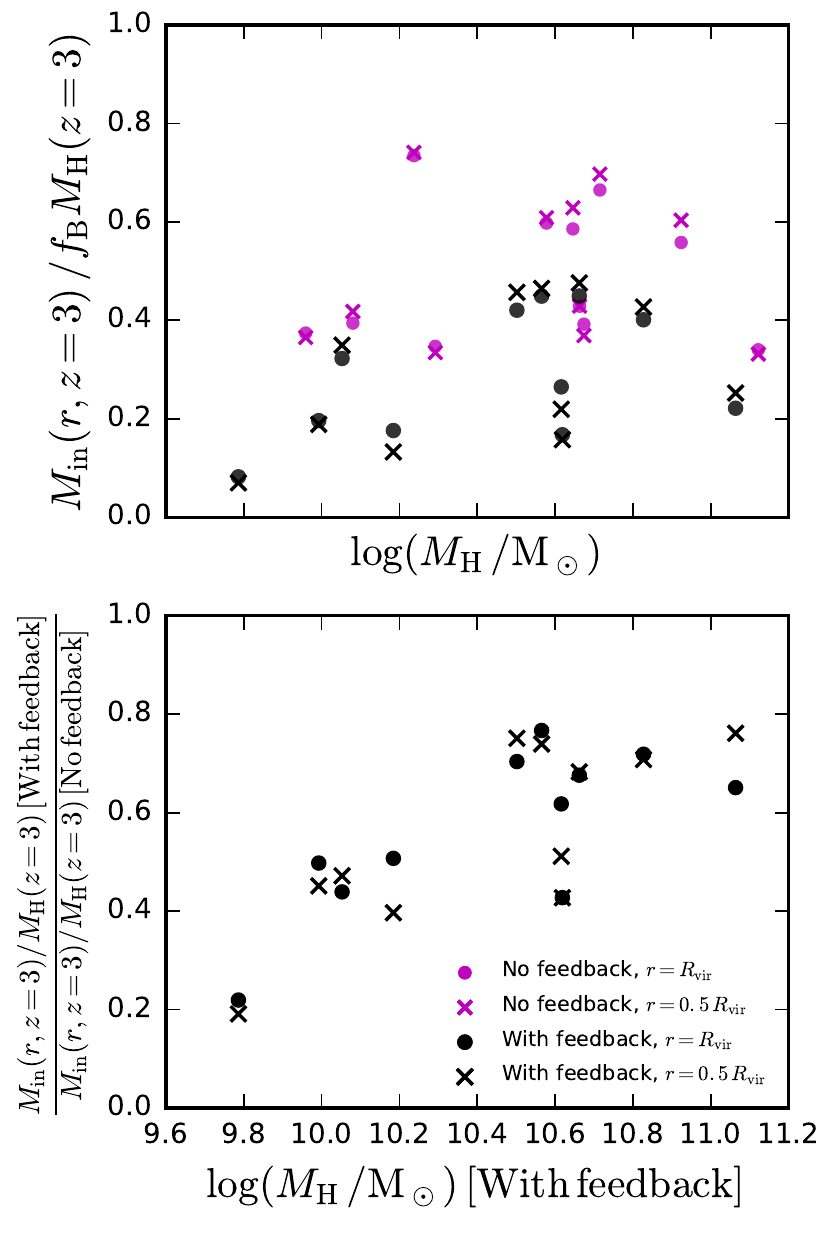}
\caption{
{\it Top:} Normalized, time-integrated inflow mass, plotted as a function of halo mass for haloes simulated with SN feedback.
The normalization factor is the baryonic mass the halo would have if it contained the universal baryon fraction.
Inflow masses are computed by integrating the inflowing mass flux from $z=8$ down to $z=3$.
Gas which is considered gravitationally bound to satellites is excluded.
Only the central main-target haloes for the $11$ zoom simulations are shown.
Star formation rates used for this quantity are computed over a $200 \, \mathrm{Myr}$ timescale.Magenta points show simulations without SN feedback. 
Black points show simulations with SN feedback.
Crosses show $M_{\mathrm{in}}(r=0.5\,R_{\mathrm{vir}})$, the inflow through a surface at half the virial radius.
Circles show $M_{\mathrm{in}}(r=R_{\mathrm{vir}})$, the inflow through a surface at the virial radius.
{\it Bottom:} The ratio of the points in the top panel, matching haloes between simulations with and without SN feedback.
}
\label{accretion_effect}
\end{figure}

\begin{figure*}
\begin{center}
\includegraphics[width=40pc]{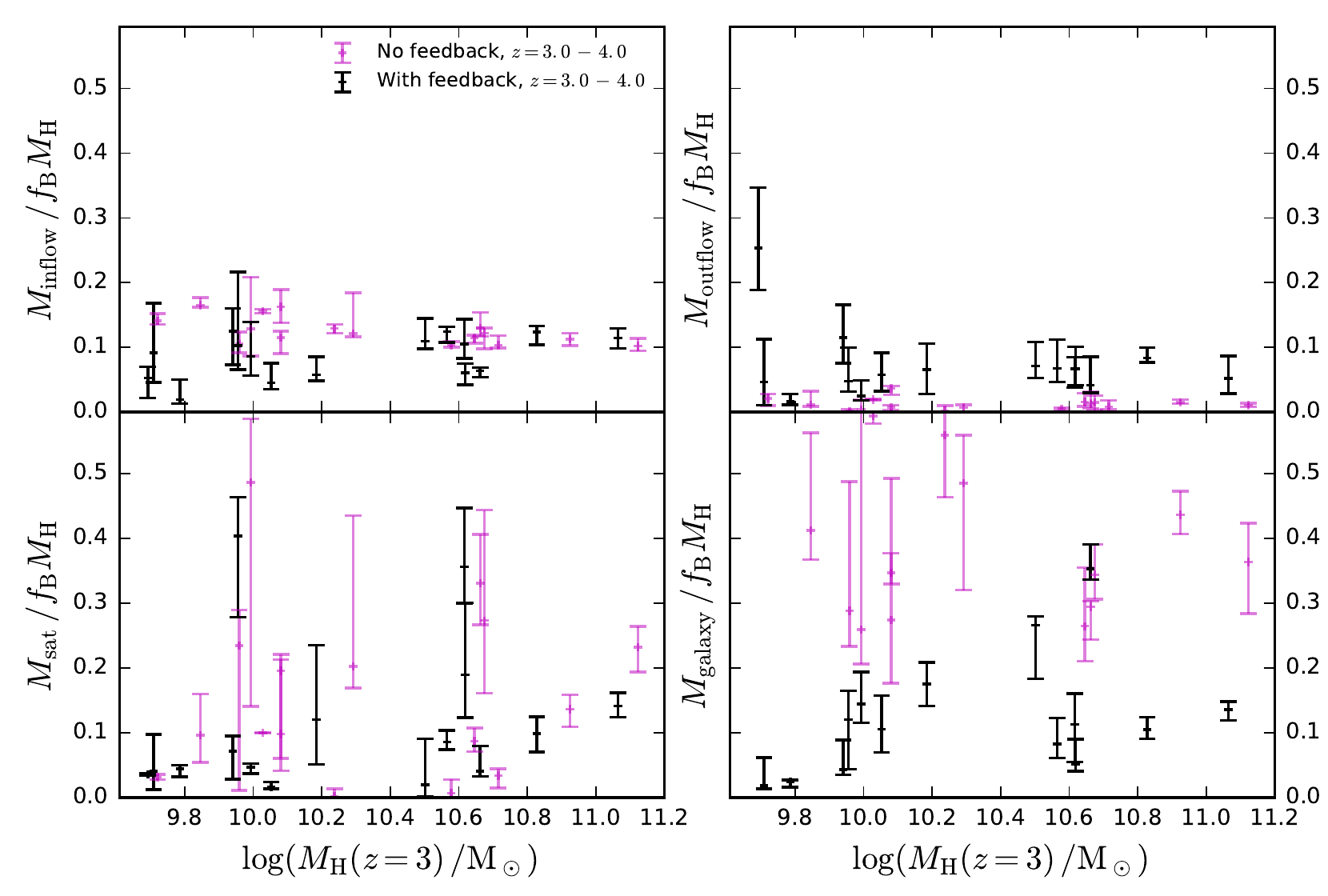}
\caption{Gas mass associated with the inflowing CGM, $M_{\mathrm{inflow}}$, the outflowing CGM, $M_{\mathrm{outflow}}$, satellites, $M_{\mathrm{sat}}$, and the central galaxy, $M_{\mathrm{galaxy}}$, plotted as a function of halo mass for central galaxies at $z=3$.
Each panel corresponds to a different baryonic component, as labelled.
Points and errorbars shows the distribution ($16^{\mathrm{th}}$, $50^{\mathrm{th}}$ and $84^{\mathrm{th}}$ percentiles) of component masses over a range of redshifts ($z=3-4$) for a single simulated halo.
Black points show haloes for simulations that include SN feedback.
Magenta points show haloes simulated without SN feedback.
The normalization factor $f_{\mathrm{B}} \, M_{\mathrm{H}}$ is the baryon mass each halo would have if it contained the universal baryon fraction.
}
\label{mcompart}
\end{center}
\end{figure*}

A second factor that can affect halo baryon fractions is the inflow rate of gas. 
Substituting the outflowing mass flux, $\dot{M}_{\mathrm{out}}(r(t'),t')$
with the inflowing mass flux, $\dot{M}_{\mathrm{in}}(r(t'),t')$, in Eqn~\ref{mout}, we define
the time-integrated  inflow mass, $M_{\mathrm{in}}$. For inflow, we choose to start the time
integral at $z=8$ because we do not have sufficient simulation outputs to accurately quantify the amount of
inflow at early times.
Fig.~\ref{accretion_effect} (top panel) shows $M_{\mathrm{in}}$, integrated from $z=8$ down to $z=3$ for two surfaces and for
simulations with and without SN feedback. Roughly speaking, the same amount of  gas that enters the virial radius also
crosses the inner surface at $0.5 \, R_{\mathrm{vir}}$, although this is complicated by gas stripping from satellites. 
We note here that the ultra-violet background does not impact the inflow of gas onto haloes in the halo mass range shown 
\cite[this effect becomes relevant for $M_{\mathrm{H}} \leq 10^9 \, \mathrm{M_\odot}$,][]{Gnedin00, Iliev07, Wise14, Onorbe17}.

Interestingly, there is a difference between the simulations with and without SN feedback. This difference is made more explicit in the bottom panel of 
Fig.~\ref{accretion_effect} which shows the ratio of points in the top panel by matching haloes between simulations with and without feedback. This shows that relative
to the no-feedback case, there can be $20-80 \%$ less  gas accretion onto, and within, haloes that are simulated with SN feedback.
The trend appears to be mass-dependent, such that inflow rates in the lower mass haloes from our sample are more affected by feedback.
This trend leads to the mass-dependent halo baryon fractions shown in Fig.~\ref{fbaryon}. The reduction in gas accretion rates due to
feedback explains $~\sim 30-50 \%$ of the total reduction in halo baryon fractions.
Note that we have normalized away the difference in final halo mass to account for the fact that in simulations with feedback, haloes are
less massive \cite[see][for a more general discussion of this effect]{Schaller15}. Without normalizing in this way, the impact of
feedback on inflow rates would be even more pronounced.

We have shown that outflows (inflows) contribute $\sim 20-60  \, \%$ ($\sim 40 \, \%$) to the reduction in halo baryon fractions.
The remaining $0-50 \, \%$ reduction in halo baryon fractions is caused by a reduction in the baryon fractions of satellites before
they are accreted onto the central haloes.

\subsection{CGM mass budget}
\label{CGM_section}

\begin{figure*}
\begin{center}
\includegraphics[width=40pc]{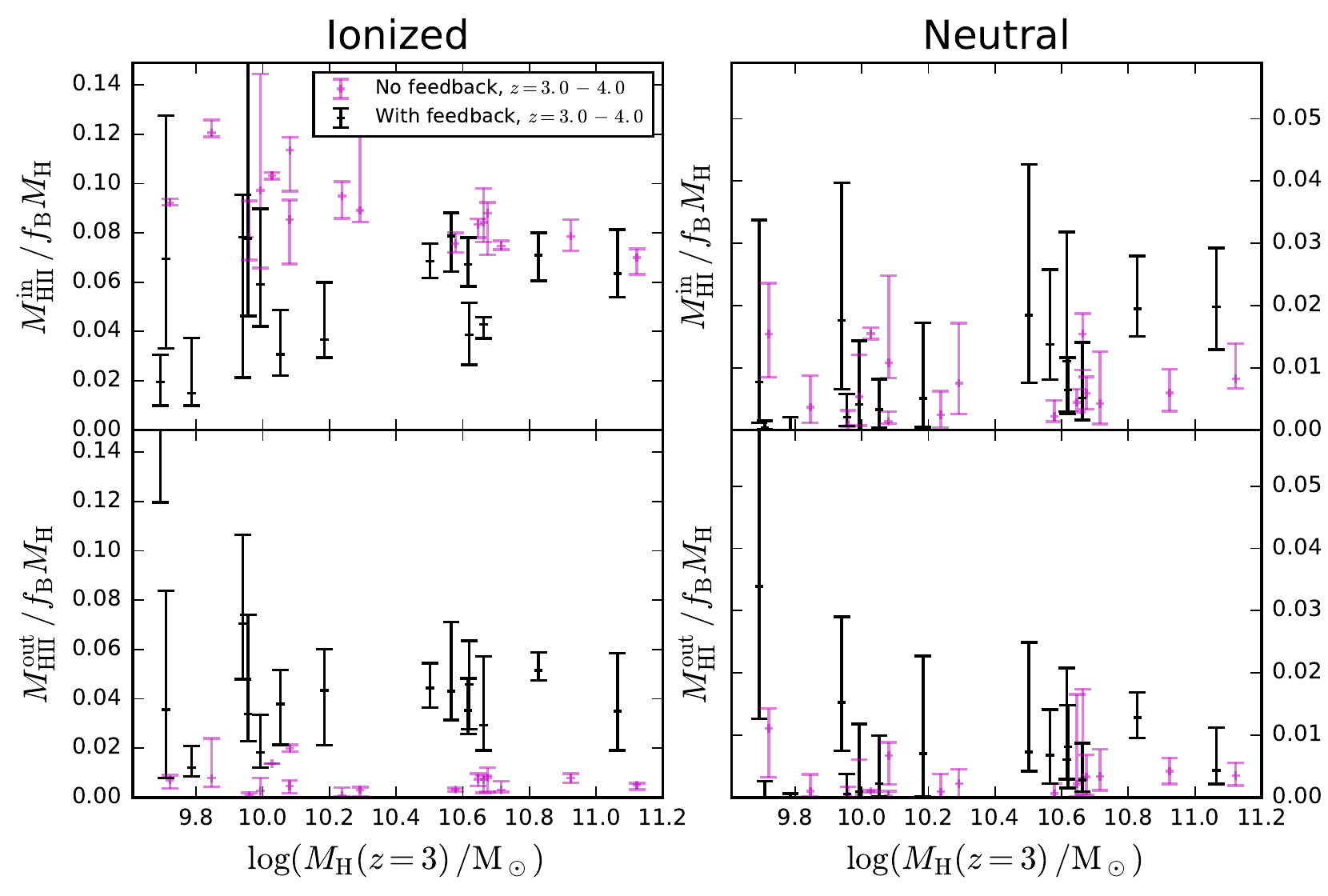}
\caption{Normalized hydrogen mass in different  circumgalactic gas components, plotted as a function of halo mass for central galaxies at $z=3$.
These components include inflowing ionized hydrogen (top-left), inflowing neutral hydrogen (top-right), outflowing ionized hydrogen (bottom-left) and outflowing neutral hydrogen (bottom-right).
Note that the dynamic range in the y-axis of the panels showing neutral hydrogen is significantly reduced compared to the panels showing ionized hydrogen.
Points and errorbars show the distributions ($16^{\mathrm{th}}$, $50^{\mathrm{th}}$ and $84^{\mathrm{th}}$ percentiles) of component masses over a range of redshifts ($z=3-4$) for a single simulated halo.
The errorbars therefore indicate the time variability in the component mass for each halo over this redshift range.
Black points show haloes for simulations that include SN feedback.
Magenta points show haloes simulated without SN feedback.
The normalization factor $f_{\mathrm{B}} \, M_{\mathrm{H}}$ is the baryon mass each halo would have if it contained the universal baryon fraction.
}
\label{mcompart_neut}
\end{center}
\end{figure*}

Following the compartmentalization scheme introduced in Section~\ref{compartment_sec},
Fig.~\ref{mcompart} compares the mass in the CGM to
the gas mass in satellites and the central galaxy. When SN feedback is included (black points), 
the mass in the  circumgalactic medium (inflows + outflows) is comparable 
to the gas mass in the central galaxy\footnote{
The two outliers (black points with 
$M_{\mathrm{galaxy}} / f_{\mathrm{B}} M_{\mathrm{H}} > 0.3$) are haloes 5 and 9 (see Table~\ref{table}), the two outliers
with unusually high stellar masses and low integrated loading factors shown in 
Fig.~\ref{mstar_mhalo} and Fig.~\ref{loading_factors}.}
The fraction of mass in satellites exhibits a substantial scatter but is typically comparable to (or greater than) the
mass in the  CGM or the central galaxy. Normalized by the final halo mass at $z=3$,
there are no obvious trends in the compartmentalization of baryons with halo mass.

While Fig.~\ref{fbaryon} shows that including supernova feedback typically reduces the net baryon
content inside haloes by half at $z=3$, Fig.~\ref{mcompart} shows that this change
is primarily associated with a reduction in the mass of the central galaxy (and from
satellites to a lesser extent). The total CGM typically contains slightly more mass 
when feedback is included, mostly because the outflowing component is significantly 
enhanced. The increase in the mass of
outflows is still modest compared to the decrease in the mass of galaxies. This
is because most of the outflowing gas moves with very high velocities, quickly leaving
the halo.

Interestingly, the reduced accretion rates discussed in Section~\ref{inflow_sne_sec} do not
translate into a significantly reduced mass of inflowing gas in the CGM. Instead, SN feedback 
reduces inflow rates primarily by reducing the radial velocity of inflowing warm/hot ionized 
gas by up to a factor two (not shown). As such, less mass is accreted onto the CGM but the gas
which is accreted spends a longer time (on average) in the CGM before infalling
onto the central galaxy. The average radial velocity of the neutral inflowing phase 
is also reduced but by a lesser amount.

\vspace{0.5cm} From an observational perspective, we are interested in
understanding how the {\it neutral} CGM is affected by feedback, as
this is the gas traced by \lya, both in absorption and emission.
Fig.~\ref{mcompart_neut} shows how the mass of hydrogen in the CGM is
divided between inflowing/outflowing and neutral/ionized phases for
redshifts in the range $3<z<4$. The mass in ionized hydrogen (left
panels) is typically a factor five larger than the mass in neutral
hydrogen (right panels) for this redshift range.  Also, the inflowing
mass is about a factor two larger than the outflowing mass for both
neutral and ionized phases, even when feedback is included. These
results do not appear to depend on halo mass.

Looking at the right panels of Fig.~\ref{mcompart_neut}, we see that
SN feedback in our simulations does not significantly increase the mass of
outflowing HI in the CGM. The increase in the outflowing mass of the CGM is instead
almost entirely due to an increase in the mass in ionized outflow.
SN feedback does increase the mass of inflowing HI by a factor $\sim 2$.
Put together, this suggests that the effect of feedback should be
visible through HI emission or absorption. However, our results
suggest that most of the signal will come from inflowing gas, which makes a
quantitative interpretation of such observables extremely challenging.

The origin of the enhanced inflowing HI content is not clear. It may
reflect a conversion of inflowing ionized hydrogen into the neutral phase.
Alternatively, if feedback slows the inflowing neutral gas without changing
the accretion rate through the virial radius, the mass of this CGM component 
will be increased. As discussed previously, we indeed find that the average
radial velocity of inflowing neutral gas is reduced slightly by feedback.

\section{Radial profiles and gas flows in the CGM}
\label{flows_sec}

\begin{figure}
\includegraphics[width=20pc]{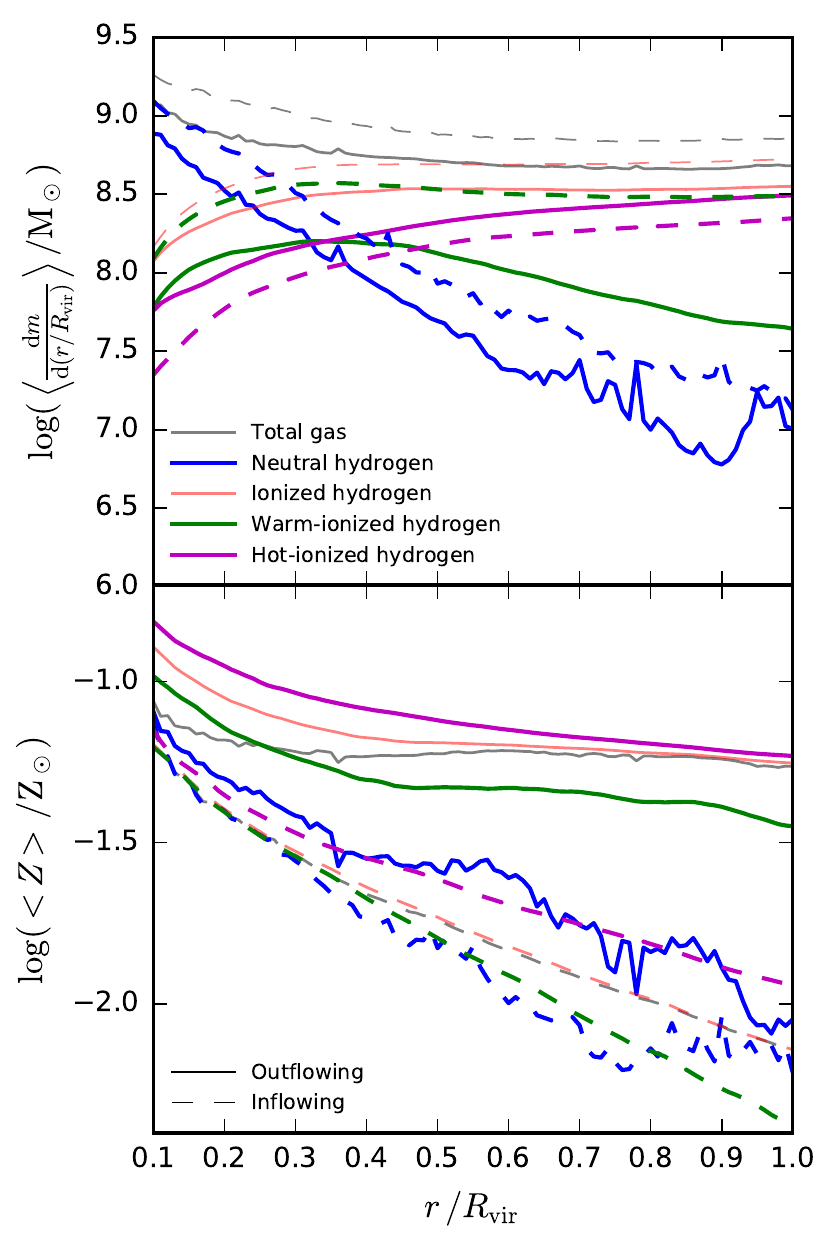}
\caption{Spherically averaged radial profiles of mass (top) and metallicity (metallicity).
Solid lines show the profiles for gas that is radially outflowing.
Dashed lines show the profiles for gas that is radially inflowing.
Black lines show the total gas, red lines show ionized hydrogen and blue lines show neutral hydrogen.
Green and magenta lines show ionized hydrogen subdivided between warm and hot phases, separated at $T=10^{4.5}\,\mathrm{K}$.
The profiles are computed by taking the mean over seven haloes, simulated with SN feedback, with mass $10.5<\log(M_{\mathrm{H}}/\mathrm{M_\odot})<11.2$, 
using all of the available simulation outputs ($15 \, \mathrm{Myr}$ spacing) over the redshift range, $3<z<4$. 
This redshift interval corresponds to a time interval of $\Delta t \sim 600 \, Myr$.
}
\label{radial_profiles_fid}
\end{figure}

Here, we explore the radial dependence of various CGM properties by stacking haloes in 
the halo mass range, $10.5<\log(M_{\mathrm{H}}/\mathrm{M_\odot})<11.2$, over the redshift
interval, $3<z<4$ (we explore other mass/redshift ranges in Appendix~\ref{ap:mz_dep}). 
Stacking yields a complete statistical description of how inflows and outflows cross the CGM, 
averaging over the strongly fluctuating inflow and outflow rates at a given radius (see Fig.~\ref{sfh_illustration}).
This allows us to address questions such as whether outflows in a given gas phase conserve mass 
as they propagate radially outwards (without needing the Lagrangian information provided by 
particle-based hydrodynamic codes or tracer particles).
Haloes are mean stacked, changing the radius variable from $r$ to 
$r/R_{\mathrm{vir}}$ and weighting simulation outputs such that haloes with different
numbers of simulation outputs contribute equally to the stack. 
Our stacking window ($600 \, \mathrm{Myr}$) is sufficiently large as to capture at least a single
single crossing of any of the gas components which we consider.

Fig.~\ref{radial_profiles_fid} 
shows the mean stacked, spherically averaged radial profiles of mass and metallicity 
(defined as the ratio of metal to total gas radial mass profiles). While satellites are 
removed, we do not attempt to subtract material from the central galaxy from these profiles. 
Roughly speaking, the cold interstellar medium of the central galaxy can extend out to 
$\sim 0.2 \, R_{\mathrm{vir}} $, and so the profiles out to these radii should be interpreted 
accordingly.

Comparing to the integrated masses presented in Fig.~\ref{mcompart} and Fig.~\ref{mcompart_neut}, 
several of the same features are apparent in Fig.~\ref{radial_profiles_fid}. There is systematically 
more mass in inflows than in outflows at all the radii considered. This is also true for the
neutral and warm-ionized gas phases (blue and green lines respectively). The hot-ionized phase
contains more mass in outflow than in inflow. Comparing the mass between the different phases,
ionized hydrogen (warm plus hot) in the CGM dominates over neutral hydrogen away from the central 
galaxy ($r \gtrsim 0.2 \, R_{\mathrm{vir}}$). Away from the central galaxy, hot-ionized hydrogen 
dominates the outflowing mass. For the inflowing CGM however, the cooler warm-ionized phase contains 
the majority of the mass, particular in the inner CGM ($r < 0.6 \, R_{\mathrm{vir}}$).

Alongside the radial mass profile, the metallicity profile contains highly complementary information.
Outflowing gas has systematically higher metallicity than inflowing gas for all of the phases shown. 
Among the various outflowing phases, hot-ionized hydrogen is the most metal rich and neutral hydrogen 
is comparatively metal poor. The total metallicity of outflow (grey solid line) is fairly constant
with radius whereas the metallicity of inflow (dashed grey line) drops by over an order of magnitude
from $0.1 \, R_{\mathrm{vir}}$ to $R_{\mathrm{vir}}$.

\begin{figure}
\includegraphics[width=20pc]{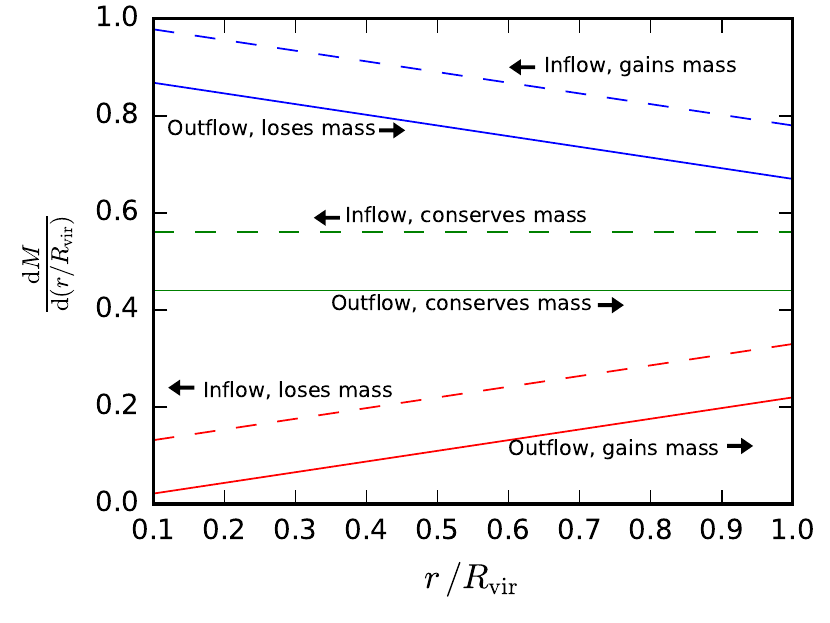}
\caption{Cartoon schematic demonstrating the relationship between radial mass profiles and mass conservation.
If inflowing (dashed lines) and outflowing (solid lines) components move with a constant radial velocity, the gradients of the radial
mass profiles can be interpreted as indicators of mass conservation.
If no mass exchange between inflow and outflow occurs, mass is conserved and the gradients of the radial mass profiles are zero (green lines).
If outflow net loses mass to inflow as it propagates outwards, the radial mass profiles have negative gradients (blue lines).
Conversely, if inflow net loses mass to outflow as it propagates inwards, the radial mass profiles have positive gradients (red lines).
Note that this schematic does not consider separate gas phases. 
For an individual gas phase, gradients in the radial mass profile depend on the rate of mass exchange between both inflowing and outflowing phases.
}
\label{cartoon}
\end{figure}

\begin{figure*}
\begin{center}
\includegraphics[width=40pc]{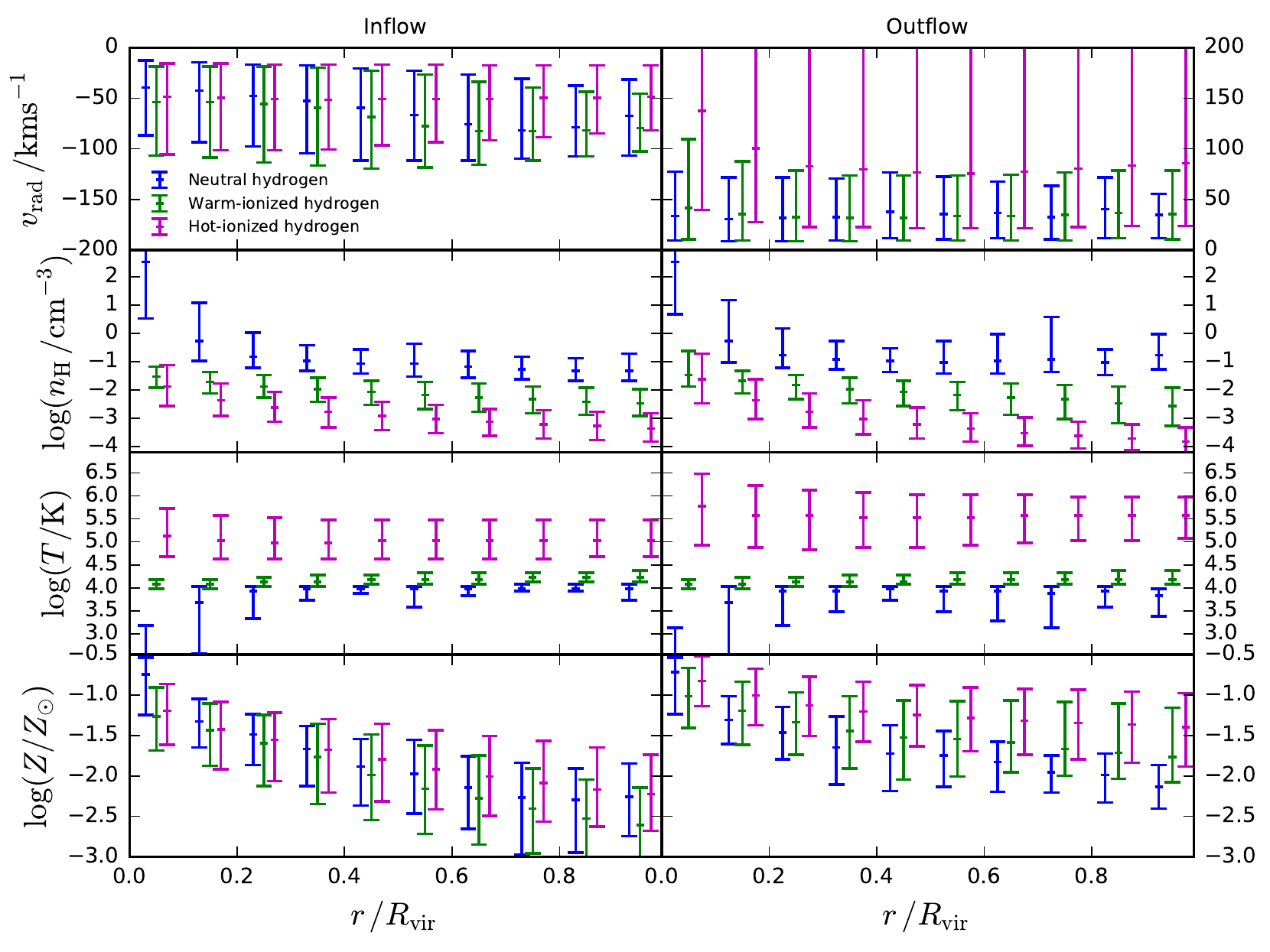}
\caption{Distributions of local gas properties plotted as a function of radius, computed by stacking haloes, 
simulated with SN feedback, with mass $10.5<\log(M_{\mathrm{H}}(z=3)/\mathrm{M_\odot})<11.2$ over the range $3<z<4$.
The left and right columns shows distributions for inflowing and outflowing gas respectively.
Gas properties shown include radial velocity (top), density (second row), temperature (third row) and metallicity (bottom).
Gas cells considered gravitationally bound to satellites are not included.
The blue, green and magenta points show the mass-weighted $16$, $50$, and $84^{\mathrm{th}}$ percentiles of the distributions
for neutral, warm-ionized and hot-ionized phases of hydrogen respectively.
}
\label{local_profiles}
\end{center}
\end{figure*}

In order to interpret further the information contained in the radial profiles shown in Fig.~\ref{radial_profiles_fid},
we note the two following points. Firstly, the CGM in our simulations is not in a quasi-hydrostatic configuration.
Rather, all of the gas phases shown exhibit significant radial velocities. The radial mass profiles should thus
be interpreted in terms of mass fluxes rather than as an equilibrium solution. Secondly, 
the radial velocity gradients of different phases are too small to explain several of the features present in the
radial mass profiles of Fig.~\ref{radial_profiles_fid} (see Fig.~\ref{local_profiles}). Instead, to 
zeroth order these radial mass profiles can be interpreted assuming the gas moves with constant radial velocities\footnote{
Note that formally written as a continuity equation the appropriate radial
profile to consider for mass conservation is the radial momentum profile (which factors in changes of the
gas radial velocity). Qualitatively, the radial momentum (not shown here) and radial mass profiles shown
in Fig.~\ref{radial_profiles_fid} are identical. This confirms that velocity changes indeed play a minor
role in shaping the radial mass profiles.}.

Fig.~\ref{cartoon} shows a cartoon which guides the interpretation of Fig.~\ref{radial_profiles_fid}.
A zero gradient for the sum of inflowing (or outflowing) phases (grey lines in Fig.~\ref{radial_profiles_fid})
implies that there is zero net mass exchange between inflowing and outflowing components, which is compatible
with a scenario where gas elements cross the entire halo without reversing direction.
Non-zero but equal gradients for outflow and inflow imply
turnaround, such that a fraction of gas elements reverse direction over the stacking time window. 
Negative radial gradients indicate mass leaving the galaxy that falls back at a later time. 
Correspondingly, positive radial gradients indicate mass that inflows through the virial radius but is ejected 
from the halo at a later point.

Fig.~\ref{radial_profiles_fid} shows that the radial mass profiles of both inflowing and 
outflowing gas (grey solid and dashed lines) are flat for $0.6<r/R_{\mathrm{Vir}}<1$ but 
show negative gradients over the radial range, $0<r/R_{\mathrm{Vir}}<0.6$. We have verified 
that these radial gradients ($\frac{\mathrm{d}}{\mathrm{d}(r/R)} \, \frac{\mathrm{d}m}{\mathrm{d}(r/R)}$) 
are almost equal for the inflowing and outflowing 
components. This therefore implies that some fraction of the mass which outflows from the halo centre turns 
around inside $0.6 R_{\mathrm{Vir}}$ and joins the inflowing component. 
This picture is supported by the metallicity profile shown in the lower panel of Fig.~\ref{radial_profiles_fid}. 
The metallicity of inflowing gas entering the halo is significantly lower than the metallicity of outflowing
gas leaving the galaxy. As such, the large negative radial metallicity gradient for inflowing gas
indicates that there must be exchanges between the outflowing and inflowing components\footnote{
Note that the total radial metallicity profile of outflowing gas (solid grey line in Fig.~\ref{radial_profiles_fid})
is flat because of the combination of metallicity segregation between different outflowing gas
phases and the changing relative mass fractions of those phases as a function of radius.}.

Focussing on the neutral phase of hydrogen (blue lines), Fig.~\ref{radial_profiles_fid} shows a strong
negative gradient for the mass profile over the entire radial range considered. This means that
outflowing neutral hydrogen is not conserved in mass as it flows outwards through the halo.
The inflowing neutral phase shows a corresponding negative gradient. A straightforward explanation for
this is that a fraction of the neutral outflow loses radial kinetic energy climbing
out of the gravitational potential, rejoining the inflowing component in a galactic fountain process.
Another possibility is that mass exchange occurs between the neutral outflow and other ionized phases (either
inflowing or outflowing), such that the neutral material is lost due to photo/collisional ionization.

Turning our attention to the inflowing phases, Fig.~\ref{radial_profiles_fid} shows that mass is continually 
lost from the hot-ionized phase as it flows further into the halo, the warm-ionized gas is approximately 
conserved and the neutral phase continually gains mass. Again, this can be contributed to by both mass exchanges 
between outflowing and inflowing components as well as mass exchanges between different phases within the
same component. 

Put together, we find that no inflowing/outflowing phase is mass conserving: none of the curves in the
upper panel of Fig.~\ref{radial_profiles_fid} are flat. Moreover, different gas phases exhibit very
different behaviour with respect to each other. If these results are representative of real galaxies,
the inference of outflow properties from absorption line measurements (both down-the-barrel and from
background quasars) becomes challenging. Specifically, there is a common assumption that outflows in
a given phase conserve mass \cite[e.g.][]{Bouche12,Heckman17}, although see \cite{Chisholm17} for a counter-example.

\subsection{Local gas properties}
\label{local_subsec}

Fig.~\ref{local_profiles} shows the local properties of the CGM, divided into different phases of 
inflowing/outflowing hydrogen. Histograms and percentiles are again computed by stacking haloes
in the same manner as for Fig.~\ref{radial_profiles_fid}. Fig.~\ref{local_profiles} shows the
inflowing gas does not exhibit evidence of any radial acceleration as it falls further into
the halo. Rather, the inflowing warm-ionized and neutral phases radially decelerate slightly for 
$r<0.5 R_{\mathrm{vir}}$ (note that the tangential velocity increases accordingly). 
Outflowing neutral and warm-ionized hydrogen moves with fairly modest radial velocities, independent 
of radius. On average, hot-ionized outflowing material moves much faster by comparison and with a 
much broader radial velocity distribution.

For both inflowing and outflowing components, warm-ionized, hot-ionized and neutral phases of hydrogen clearly
segregate in density and temperature (by construction for the hot/warm divide). Comparing neutral inflow and outflow,
it is interesting that neutral outflowing hydrogen has broader density and temperature distributions, indicating 
that part of this phase is more associated with clumps while neutral inflow is more diffuse on average. This can be
seen visually in Fig.~\ref{density_maps}. This difference between neutral inflow and outflow disappears at higher
redshifts ($z \geq 4$, not shown), with the neutral outflow instead showing a narrower density distribution similar to the
inflowing neutral hydrogen density distribution shown in Fig.~\ref{local_profiles} for lower redshift.

The metallicities of different inflowing phases are fairly similar to each other while the metallicities 
of outflowing phases are more segregated. For outflowing gas the level of metallicity 
segregation between different phases does not decrease with increasing radius (rather the opposite), 
implying that different outflowing phases do not significantly mix with each other as they propagate outwards.
For inflowing gas, metallicities rise strongly as gas moves towards the centre. This may be
the result of a galactic fountain in which enriched outflowing gas falls back onto the galaxy.
Alternatively, gradual metal enrichment of the inflowing gas occurs through mixing with
the enriched outflowing phases. In practice both processes presumably occur.

\subsection{Energy decomposition of inflow/outflow}
\label{energy_subsec}

\begin{figure}
\includegraphics[width=20pc]{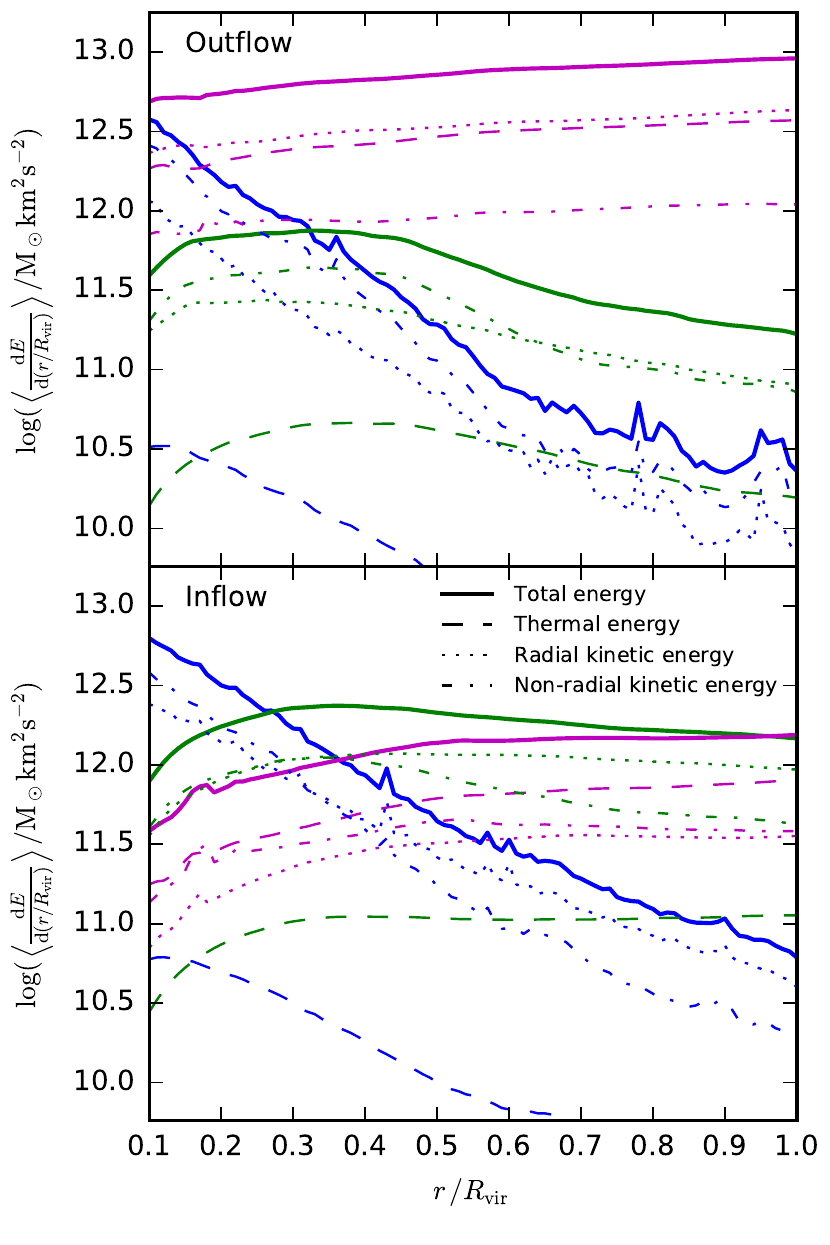}
\caption{Spherically averaged radial profiles of energy, plotted as a function of radius inside haloes.
The profiles are separated between radially outflowing (top panel) and inflowing (bottom panel) gas.
The profiles are further separated between neutral (blue lines), warm-ionized (green) and hot-ionized (magenta) phases of hydrogen. 
Solid lines show total energy (thermal plus kinetic).
Dashed lines show thermal energy.
Dotted lines show kinetic energy associated with radial motion.
Dash-dotted lines show the remaining kinetic energy associated with non-radial motion.
The halo sample and stacking time interval are the same as \protect Fig.~\ref{radial_profiles_fid}.
}
\label{radial_profiles_energy}
\end{figure}

Additional information on the nature of gas flows in our simulated haloes is shown in 
Fig.~\ref{radial_profiles_energy}, which shows the decomposition between radial-kinetic,
tangential-kinetic, and thermal energy for the same halo stack as in Fig.~\ref{radial_profiles_fid}. 
Outflowing hot-ionized hydrogen dominates in energy
and is clearly comprised of a coherent, thermalized radial outflow. In contrast, the warm-ionized
outflow contains negligible thermal energy and a comparable amount of radial-kinetic and
tangential-kinetic energy. The neutral outflow contains systematically more tangential-kinetic
energy for most of the radial range considered. This shows that the warm-ionized and neutral
outflowing CGM is not in a coherent radial wind, and instead exhibits a more complex velocity
structure.

Focussing instead on inflows, Fig.~\ref{radial_profiles_energy} shows that as inflowing 
gas approaches the halo centre, radial-kinetic energy is transferred to tangential kinetic energy. For the
neutral and warm-ionized phases, tangential kinetic energy is the largest contributor
to the total energy of these inflowing phases in the inner regions ($r<0.4 \, R_{\mathrm{Vir}}$).
This is consistent with the slight radial deceleration seen in Fig.~\ref{local_profiles}
for inflowing warm-ionized and neutral hydrogen in this radial range.

\subsection{How does SN feedback shape radial mass flows in the CGM?}
\label{fb_effect_section}

\begin{figure*}
\begin{center}
\includegraphics[width=40pc]{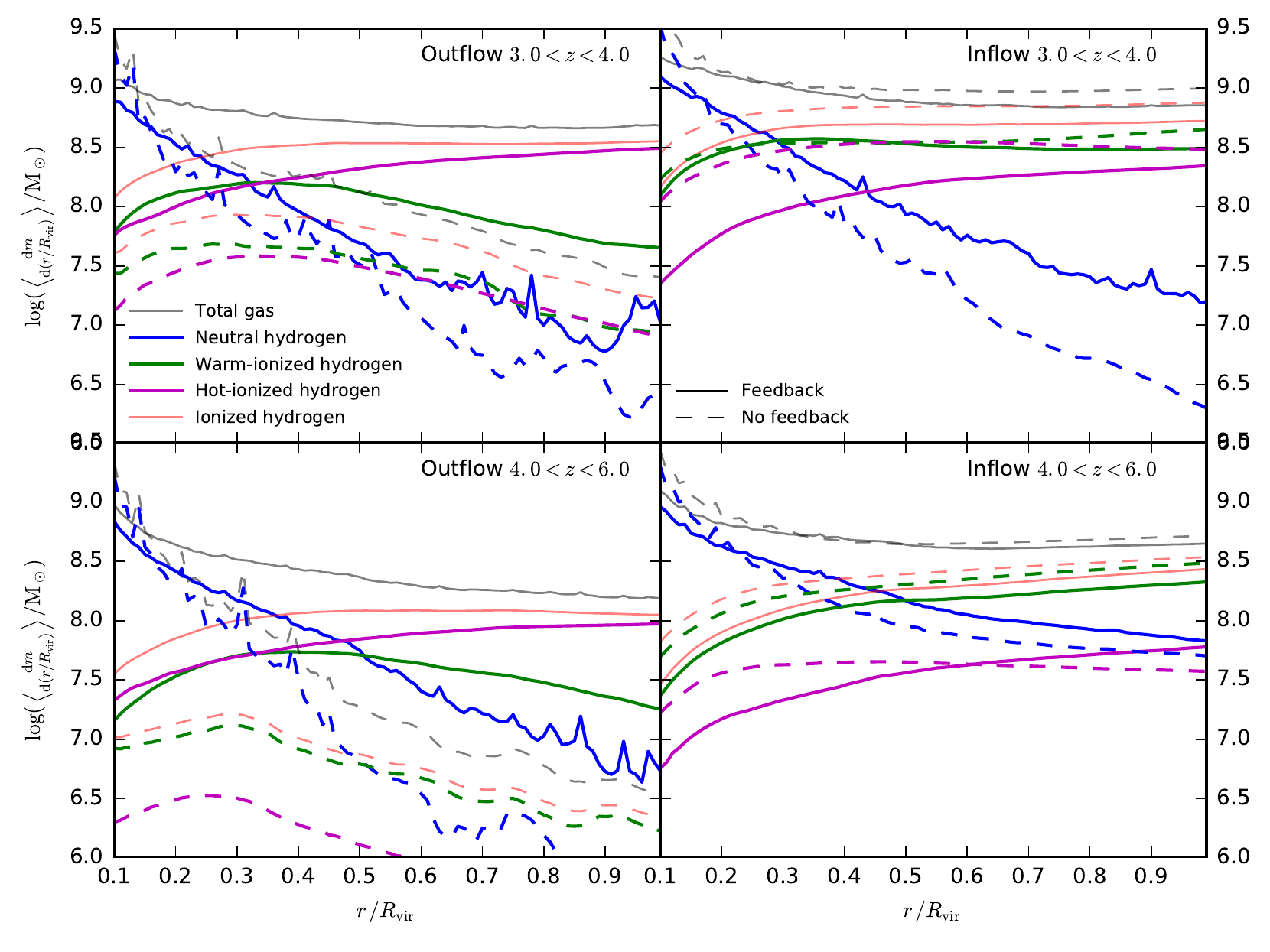}
\caption{The effect of supernova feedback on mean stacked, spherically averaged radial mass profiles in the CGM.
Solid lines correspond to simulations that include feedback and dashed lines correspond to simulations that do not include feedback (note the formatting change from \protect Fig.~\ref{radial_profiles_fid}).
The left/right panels show profiles for gas that is radially outflowing/inflowing respectively.
Different line colours correspond to different combinations of warm-ionized, hot-ionized and neutral hydrogen, as labelled.
Black lines show the total gas profile (including helium).
The top/bottom panels correspond to different redshift intervals, as labelled.
As in \protect Fig.~\ref{radial_profiles_fid}, the haloes included in the stack have $10.5<\log(M_{\mathrm{H}}(z=3)/\mathrm{M_\odot})<11.2$.
}
\label{radial_profiles_nfb}
\end{center}
\end{figure*}

Expanding on the spatially integrated analysis presented in Section~\ref{CGM_section}, 
the impact of SN feedback on the CGM is shown in more detail
in Fig.~\ref{radial_profiles_nfb}, which compares stacked radial mass profiles
between simulations with and without feedback. Consistent with Fig.~\ref{mcompart_neut},
over the redshift interval, $3<z<4$, the stacked neutral hydrogen mass increases
in the CGM at all radii when feedback is included (aside from close to the halo 
centre). 

For outflowing neutral hydrogen, the enhancement is modest, showing supernova
feedback does not significantly alter the radial mass distribution in this component for this redshift/mass
interval \footnote{Note that for simulations without feedback, outflowing material stripped
from satellites and  material moving past pericentric passage can still form
an outflowing component.}. On the other hand, the high-density ($n_{\mathrm{H}} > 1 \, \mathrm{cm^{-3}}$) 
and low-temperature ($T < 4000 \, \mathrm{K}$) part of the outflowing neutral component 
seen in Fig.~\ref{local_profiles} is only seen when SN feedback is included.
This implies that SN feedback does result in a clumpier outflowing neutral
CGM.

Over the higher redshift interval ($4<z<6$) shown in the bottom panels of
Fig.~\ref{radial_profiles_nfb}, the inclusion of feedback has a much
more significant effect in enhancing the mass of neutral outflowing hydrogen.
The mass in inflowing neutral hydrogen is also enhanced by feedback (both relatively and absolutely)
in the higher redshift interval, as in the $3<z<4$ interval. Unlike
the outflow case, neutral inflow is more significantly enhanced in the
lower redshift interval.

\section{Discussion}
\label{discussion_section}

\subsection{Implications for spatially-extended \lya emission}

In future work, we plan to utilize the simulations presented here coupled with radiative
transfer calculations to explore the implications
of the recently observed, spatially-extended \lya emission around individual faint, high
redshift galaxies \cite[][]{Wisotzki16}. A key application of such observations is that
they are presumably sensitive (at some level) to the details of SN feedback, offering
an alternative to analyses of HI covering fractions inferred using absorption from background
quasars (which suffer from limited statistics and the difficulty of linking absorbers
to galaxies).

Here, we have demonstrated that the inclusion of
supernova feedback both significantly reduces the gas content of simulated central galaxies
and increases the neutral CGM content. Both these effects are expected to enhance
the spatially extended \lya signal. Decreasing the neutral gas content at the centre 
of the halo is expected to increase the escape fraction of \lya photons emitted around
young stars in the central galaxy, increasing the number of photons that can be scattered
from CGM gas. A competing effect however is the reduction of star formation associated with
the inclusion of feedback. Enhanced neutral gas content in the CGM will both increase the
amount of direct \lya emission from the CGM (cooling radiation, fluorescence) and lead
to increased scattering further away from any central source of \lya photons.

Fig.~\ref{density_maps} illustrates the typical projected spatial configuration of neutral
inflowing and outflowing gas. Inflowing gas is anisotropically distributed along filamentary
structures, containing the majority of the neutral hydrogen mass. Outflowing material,
while subdominant in mass, has a tendency to be aligned perpendicular to the direction of inflow and
can act to fill the projected surface with neutral hydrogen, potentially forming an important
contributor to the scattered \lya signal. The effect of feedback is less pronounced
on the neutral outflow for $z=3-4$. This emphasises the importance of satellites (and their associated tidal material) 
to the neutral outflowing CGM content. The high mass resolution of our simulations makes
them ideal for capturing this effect. At higher redshifts \cite[$z=4-6$, also observed by][]{Wisotzki16},
the effect of feedback on the neutral, outflowing CGM content becomes more significant, indicating
that \lya emission may offer greater constraining power on subgrid feedback modelling
at these redshifts.

Also of interest for \lya emission is the velocity structure of the neutral inflowing/outflowing
CGM. Our simulations predict only modest radial velocities ($50-100 \, \mathrm{km s^{-1}}$) for both
inflowing and outflowing neutral phases. Within half the halo virial radius, a larger fraction of the kinetic
energy of neutral gas is in tangential motion. In particular, the outflowing neutral CGM never resembles
a coherent radial outflow.
This picture is somewhat in tension with the properties of the \lya line measured around LAEs, which 
tends to be redshifted with respect to systemic \cite[e.g.][]{Erb14,Song14} and is most simply explained as the result of back-scattering from neutral
outflow moving at larger radial velocities than those reported here \cite[][]{Hashimoto15}.
This may indicate that the properties of the neutral CGM
predicted by our numerical simulations are not accurate. Alternatively, it is also possible that
the spectral properties of observed \lya lines are imprinted on the smaller scales of the ISM.
In principle, our simulations combined with \lya radiative transfer can capture these small
scale effects and we plan to return to this question in future work.

\subsection{Numerical convergence of neutral CGM content}

A key limitation of cosmological simulations is that the spatial resolution rapidly degrades away from the high-density
ISM at the centres of haloes. In the CGM, this effect will act to smooth out density fluctuations in the gas \cite[][]{Fumagalli14}. This can
impact the neutral CGM content in a number of ways. Low density gas will be artificially enhanced in density (by mixing
with higher density gas), enabling
enhanced radiative cooling, spuriously increasing the neutral fraction. Conversely, moderate density gas which is 
self-shielded against the UV background but not dense enough to self-gravitate (and thus trigger refinement, preserving
the resolution) will be artificially mixed with lower density ambient gas. These two effects may balance to produce
the same net neutral CGM mass but will nonetheless smooth out the spatial structure, with obvious implications for \lya 
radiative transfer. In AMR codes, non-Lagrangian refinement criteria can be employed to explore the impact of these
effects. For example, the refinement strategy of \cite{Rosdahl12}, where refinement is triggered proportional to the spatial
gradient of the ionization fraction will act to prevent artificial mixing of low/moderate density gas, alleviating this
effect. We leave the application of such a refinement scheme to the simulations presented here to future work.

\subsection{Missing physics}
\label{missing_section}

For the more massive simulated haloes in our simulated sample, the stellar masses appear to be on average too high compared to extrapolations
of observational constraints (see Fig.~\ref{mstar_mhalo}). This indicates that either the efficiency of star formation or the efficiency of
feedback is over/under estimated respectively\footnote{Despite the fact that we over-inject SN energy by a factor $\approx 2-3$, see Section~\ref{sn_section}.}.  
Also shown in Fig.~\ref{mstar_mhalo} is the reference \eagle simulation, for which
the predicted masses agree very well with observational constraints at low redshift \cite[][]{Schaye15}, and for $z=3,4$ at higher masses
\cite[where observational constraints are available][]{Furlong15}. This agreement is achieved in \eagle by rigidly controlling the radiative
losses: gas is only heated by SN if the temperature difference is $\Delta T = 10^{7.5} \, \mathrm{K}$, heating gas well above the peak of the cooling curve \cite[][]{Schaye15}.
In our simulations (which have $ \sim 10^3$ higher mass resolution and allow a cold phase to form), such an approach is not
satisfactory given the ability to resolve distinct phases of SNe explosions \cite[energy conserving versus momentum conserving,][]{Kimm15}. Nonetheless, it                  
has been demonstrated in idealised simulations of galaxy disks (at a similar resolution to cosmological simulations
presented here) that the delayed-thermal feedback scheme used in \eagle gives similar results to the mechanical feedback
scheme used here (when both feedback schemes are used in \ramses at similar resolution and with the same star-formation
and cooling models) \cite[][]{Rosdahl17}.

Zoom simulations have also had success in reproducing the stellar masses inferred from observations \cite[e.g.][]{Hopkins14,Agertz15,Wang15}.
\cite{Wang15} achieve this by artificially limiting radiative losses, and by utilising very efficient radiative energy
ejection into the ISM from massive stars.
\cite{Agertz15} show that they can reproduce observational constraints by combining energy and momentum injection
from massive stars and supernova with a higher local star formation efficiency (they do not artificially prevent radiative losses).
\cite{Hopkins14} present a similar argument, emphasizing the importance of coupling between stellar winds, radiation pressure (including efficient trapping of infrared photons)
and SN explosions for regulating star formation and generating powerful mass outflows.
\cite{Muratov15} analyse the same simulations and report mass-loading factors which are comparable 
to the lower mass haloes in our sample and are on average a factor $\approx4$ higher than the more massive haloes in our sample \cite[see also][]{AnglesAlcazar16,Keller16}. Increasing
the outflow rates to such levels in our simulations for the more massive haloes would likely help to achieve
agreement with observational constraints on stellar masses. We also note that the efficient feedback implemented in these simulations
leads to a significant enhancement of neutral CGM content with respect to older simulations \cite[][]{Faucher-Giguere15}.

A common feature of the \cite{Hopkins14}, \cite{Wang15} and \cite{Agertz15} simulations is that they all
employ efficient early feedback, which can pre-process the ISM before supernovae explode \cite[indeed the important role of
these mechanisms in changing the impact of SN explosions has been explored in detail by the ISM research community, e.g.][]{Geen15,Fierlinger16,Wareing17}. As such, the absence
of these feedback mechanisms in our simulations could very plausibly explain our high predicted stellar masses, despite
the fact that we inject more SNe energy per unit stars formed than in these simulations.
It remains to be seen however if results from such cosmological simulations
can be reproduced with full, coupled radiation-hydrodynamics \cite[particularly the high assumed efficiency of radiative feedback, see][]{Rosdahl15, Peters16}.

Aside from radiation pressure and photoheating from local sources, there are other feedback mechanisms which
are entirely absent from our simulations at present. The inclusion of stellar winds has been argued by some 
authors as a crucial mechanism to regulate star formation in very high-resolution simulations 
\cite[e.g.][]{Gatto16, Peters16}. Also, the importance of cosmic rays for shaping the ISM and
driving outflows has been explored by various authors recently \cite[e.g.][]{Booth13,Salem14,Girichidis16,Simpson16,Pakmor16}.

\section{Summary}
\label{summary_section}

We have presented a sample of $11$ cosmological zoom simulations, run down to $z=3$, specifically designed to complement CGM
observations around faint, high redshift galaxies. We show that observations from \cite{Wisotzki16}
allow the neutral CGM to be probed in \lya emission out to close to the virial radii of
individual $z=3-6$, $M_{\mathrm{H}} \sim 10^{11} \, \mathrm{M_\odot}$ haloes. Haloes in this mass and
redshift range can be simulated at reasonable computational cost at high mass and spatial resolution,
opening a new window for constraining the physics of gas flows around high-redshift galaxies.

To act as a baseline for future simulations, we present here simulations including type II SN
as the sole source of feedback. This leads to stellar masses which are larger at
given halo mass than is implied by extrapolation of observational constrains. Galaxy sizes and star formation
rates are not in obvious tension with observations (again involving some extrapolation).
Put together, the low mass of our haloes precludes forming robust conclusions when comparing to current observations.

The simulations presented here are allowed to form a cold ISM phase, a crucial step for 
computing realistic \lya radiative transfer through the ISM. We also use subgrid
models for star formation and SN feedback which are designed to operate at the high
resolution used here, where the ISM is resolved well below $\mathrm{kpc}$ scales.
In this paper, we lay the groundwork for interpreting future work on \lya radiative
transfer using these simulations, exploring the neutral CGM content
of high-redshift haloes. Our main results are summarised as follows:

\begin{itemize}

\item Supernova feedback acts to reduce the integrated baryon content of simulated haloes by $40-85 \%$ (Fig.~\ref{fbaryon} in Section~\ref{integrated_section}),
partly by driving outflows (with mass-loading factors ranging between $15$ and unity, Fig.~\ref{loading_factors}) and partly by reducing gas inflow rates (by $20-80 \%$, Fig.~\ref{accretion_effect}).

\item The time-averaged mass in the inflowing CGM is roughly unaffected by SN feedback (Fig.~\ref{mcompart} in Section.~\ref{CGM_section}), but
the radial velocity of inflowing warm and hot ionized gas phases is reduced by a factor two.

\item Including SN feedback affects the phase distribution of inflowing hydrogen. The mass of ionized (neutral) inflowing
hydrogen in the CGM is reduced (increased) by $ \sim 20 \%$ (a factor $ \sim 2$, Fig.~\ref{mcompart_neut} in Section~\ref{CGM_section}).

\item The time-averaged mass in the outflowing CGM is significantly enhanced by SN feedback, mostly in the
ionized phase of hydrogen (Fig.~\ref{mcompart_neut} in Section~\ref{CGM_section}), although the neutral outflowing CGM is also significantly enhanced for $z>4$ (Fig.~\ref{radial_profiles_nfb} in Section~\ref{fb_effect_section}).

\item The mass in outflowing neutral hydrogen declines strongly as outflowing gas moves away from the halo centre (Fig.~\ref{radial_profiles_fid} in Section.~\ref{flows_sec}).
Correspondingly, the mass in inflowing neutral hydrogen increases strongly as inflowing gas moves toward the halo centre, suggestive of a galactic fountain process.

\item On the other hand, the relative mass fractions of different outflowing phases of hydrogen change as a function of radius, indicating possible mass exchange between different outflowing phases.
The same is true for the corresponding inflowing phases. 

\item For $z=3-4$, the density distribution of outflowing neutral hydrogen is broader than the density distribution of neutral inflow (Fig.~\ref{local_profiles} in Section~\ref{local_subsec}).
For $z>4$, the density distribution of outflowing neutral hydrogen instead closely resembles the density distribution of neutral inflow, indicating an evolution of the average neutral CGM properties between $z=3-4$ and $z=4-6$.

\item Neutral outflow moves with modest radial velocities ($ \sim 50 \, \mathrm{kms^{-1}}$, Fig.~\ref{local_profiles} in Section~\ref{local_subsec}), with the majority of the kinetic energy instead in tangential motion with respect to the halo centre (Fig.~\ref{radial_profiles_energy} in Section~\ref{energy_subsec}).

\end{itemize}

Put together, we demonstrate that SN feedback has a significant effect on a number of properties of the neutral/ionized CGM.
While these simulations are in tension with extrapolations to observationally inferred constraints on stellar masses, it remains to be seen if they can
reproduce the spatially extended \lya emission observed by MUSE, or the properties of LLS observed in high-redshift QSO spectra (we defer
both comparisons to future work).
Moving forwards, these observational probes of the CGM around low mass, high-redshift galaxies will provide valuable insight for 
constraining more detailed cosmological simulations that include full radiation-hydrodynamics and non-equilibrium chemistry, particularly given that
such simulations are only computationally feasible for this mass/redshift range at present.

\section*{Acknowledgements}

We thank Matthieu Schaller for making data from the no-feedback \eagle simulation available to us.
We are grateful to the LABEX Lyon Institute of Origins (ANR-10-LABX-0066) of the Universit\'{e} de Lyon for its financial support within the program ``Investissements d'Avenir'' (ANR-11-IDEX-0007) of the French government operated by the National Research Agency (ANR).
Simulations were run at the Common Computing Facility (CCF) of LABEX LIO. 
JB and JR acknowledge support for the ORAGE project from the Agence Nationale de la Recherche under grant ANR-14-CE33-0016-03.
TK acknowledges support by the ERC Advanced Grant 320596 ``The Emergence of Structure during the Epoch of Reionization” and the National Research Foundation of Korea to the Center for Galaxy Evolution Research (No. 2017R1A5A1070354).

----------------------------------------------
\bibliographystyle{mn2e}
\bibliography{bibliography}
---------------------------------------------------------------------

\appendix
\input{appendix_mz_dep}

\label{lastpage}
\end{document}

%% file: appendix_mz_dep.tex
\section{Radial mass profiles: Mass and redshift trends}
\label{ap:mz_dep}

\begin{figure*}
\begin{center}
\includegraphics[width=40pc]{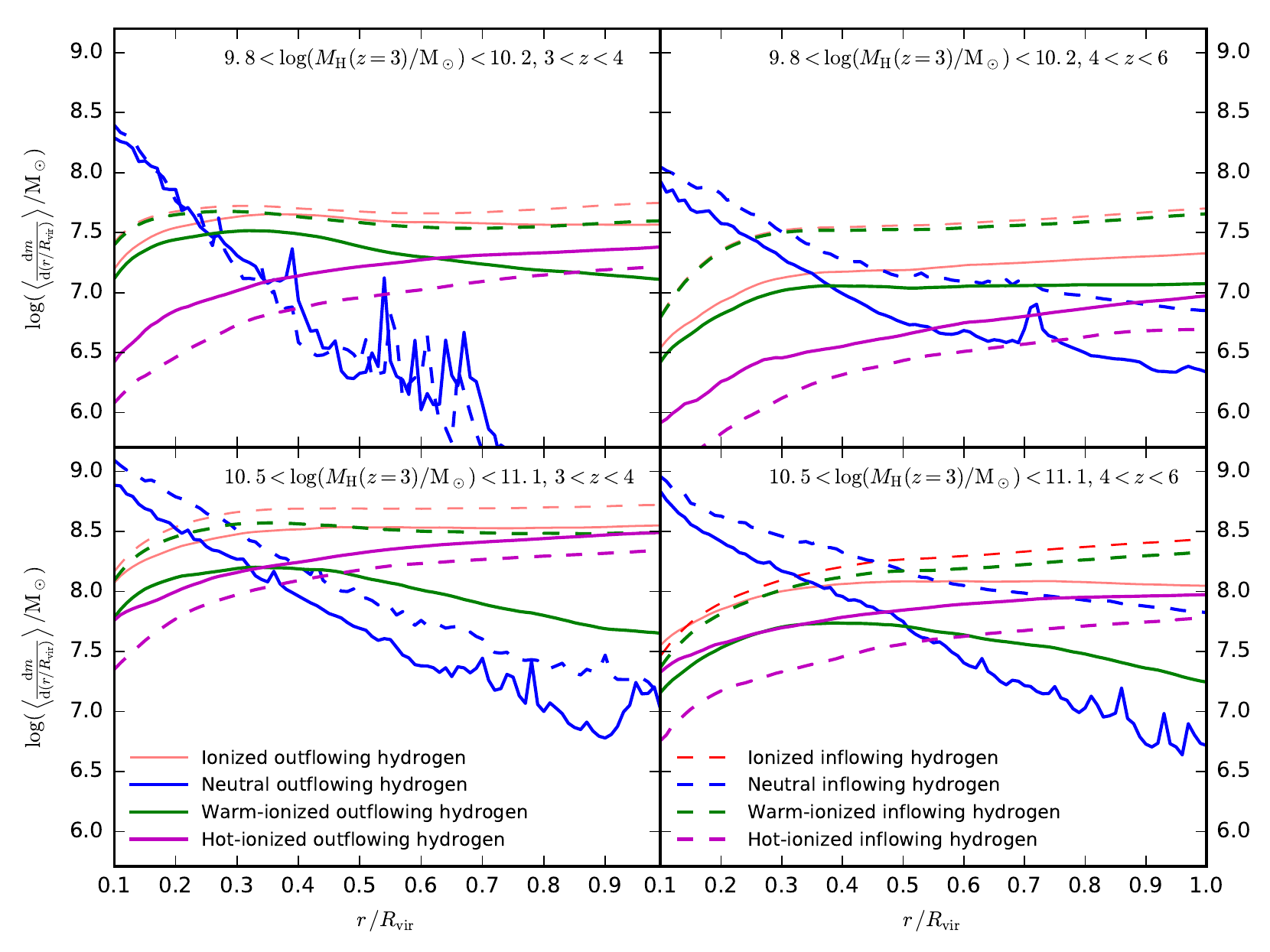}
\caption{Mass and redshift trends for different phases of  circumgalactic hydrogen.
Spherically integrated radial mass profiles are plotted as a function of radius. The
profiles are split between different combinations of inflowing/outflowing and
warm-ionized/hot-ionized/neutral phases of hydrogen.
Formatting follows \protect Fig.~\ref{radial_profiles_fid}.
Each panel corresponds to a different combination of stacking intervals in halo mass and redshift, as labelled.
}
\label{radial_profiles_m_z_dep}
\end{center}
\end{figure*}

To explore how the results presented in Fig.~\ref{radial_profiles_fid} depend on redshift and halo mass,
we present radial mass profiles for two stacking intervals in redshift and two intervals in halo mass in
Fig.~\ref{radial_profiles_m_z_dep}. While the radial gradients of each phase are qualitatively
similar for each of the stacks considered, the relative contributions in mass from each phase
do vary significantly from stack to stack.

Focussing on neutral CGM content, in general, the relative
importance of the neutral CGM increases with redshift. In the lower halo mass interval, the
relative contribution from both neutral inflow and outflow are significantly enhanced
in the higher redshift interval. For the higher halo mass interval, the relative contribution of
neutral inflow is significantly enhanced in the higher redshift interval but neutral outflowing mass
remains comparable with redshift.

For the ionized phases, the relative contribution of hot-ionized inflow drops significantly
at higher redshifts while warm-ionized hydrogen is always the dominant inflowing phase
(by mass) away from the central galaxy ($r>0.4 R_{\mathrm{vir}}$). For outflowing ionized
phases, the hot phase always dominates for the more massive halo stack. For the lower mass
stack, the warm-ionized phase is comparable to the hot-ionized phase, and forms the largest
contributor to outflowing mass at all radii in the $4<z<6$ stack.